\documentclass[12pt]{article}
\usepackage{putex}
\usepackage{graphicx}

\providecommand{\abs}[1]{\lvert#1\rvert}

\begin{document}

\preprint{hep-th/0703090 \\ PUPT-2227}

\institution{PU}{Joseph Henry Laboratories, Princeton University, Princeton, NJ 08544}

\title{Master field treatment of metric perturbations sourced by the trailing string}

\authors{Steven S.~Gubser and Silviu S.~Pufu}

\abstract{We present decoupled, separable forms of the linearized Einstein equations sourced by a string trailing behind an external quark moving through a thermal state of ${\cal N}=4$ super-Yang-Mills theory.  We solve these equations in the approximations of large and small wave-numbers.}

\date{March 2007}

\maketitle
\tableofcontents

\section{Introduction}
\label{INTRODUCION}

In \cite{Friess:2006fk}, using the AdS/CFT correspondence \cite{Maldacena:1997re,Gubser:1998bc,Witten:1998qj} (for a review see \cite{Aharony:1999ti}), the expectation value $\langle T_{mn} \rangle$ of the stress tensor of ${\cal N}=4$ super-Yang-Mills theory at finite temperature was calculated in the presence of an external quark moving a constant speed $v$.  The calculation is based on the trailing string configuration of \cite{Herzog:2006gh,Gubser:2006bz} (see also \cite{Casalderrey-Solana:2006rq} for a closely related development, \cite{Sin:2004yx} for earlier work in a somewhat similar vein, and, for example, \cite{Gubser:2006nz} for a brief survey of other recent literature on related topics).  Conceptually, all that one requires in order to calculate $\langle T_{mn} \rangle$ is a solution of the linearized Einstein equations in the $AdS_5$-Schwarzschild background, sourced by the trailing string, with infalling boundary conditions imposed at the horizon and additional boundary conditions, appropriate to the absence of deformations of the lagrangian of ${\cal N}=4$ gauge theory, imposed at the boundary of $AdS_5$-Schwarzschild.  In practice, the calculation is somewhat tedious because there are fifteen linearized Einstein equations.  In \cite{Friess:2006fk} these equations were somewhat streamlined from their original form in ``axial'' gauge, where all metric perturbations with an index in the direction orthogonal to the boundary are set to zero.  However, the most interesting set of equations in \cite{Friess:2006fk}, namely the one that determines $\langle T_{00} \rangle$, still involves four coupled second order differential equations for four unknown functions, together with three first order constraints.  The complicated form of these equations makes them hard to work with, challenging to check independently, and difficult to extend to more general settings---for example, backgrounds where conformal invariance is broken or the motion of the quark is not uniform.

There exists in the GR literature a highly developed ``master field'' formalism, dating back to such works as \cite{Regge57, Zerilli:1970se,Teukolsky:1973ha}, whose aim is to obtain fully decoupled and separable forms of the linearized Einstein equations, as well equations for other fields with spin, in curved backgrounds with some symmetry.  As a toy version of the calculation, consider a massive scalar, with action
 \eqn{ScalarAction}{
  S = \int d^5 x \,\sqrt{-g} \left[ -{1 \over 2} (\partial\phi)^2
   - {1 \over 2} m^2 \phi^2 + J\phi \right] \,,
 }
in the curved background
 \eqn{CurvedBackground}{
  ds^2 = a(r)^2 \left[ -h(r) dt^2 + d\vec{x}^2 \right] +
    {dr^2 \over h(r) a(r)^2} \,.
 }
The first step is to consider a specific Fourier mode of $\phi$ and $J$ in the $\vec{x}$ directions, with wave-number $\vec{k}$.  The equation of motion for this mode is
 \eqn{WaveEqnK}{
  \left[ -{1 \over f} \partial_t^2 + {1 \over a^3}
   \partial_r a^3 f \partial_r -
   {k^2 \over a^2} - m^2 \right] \phi = -J \,,
 }
where $k=|\vec{k}|$ and we have defined $f(r) = h(r) a(r)^2$.  The corresponding master field and master equation are an equivalent way of writing \eno{WaveEqnK}:
 \eqn[c]{MassiveMaster}{
  (\square_2 - V_m) \Phi_m =
   \left[ -{1 \over f} \partial_t^2 + \partial_r f \partial_r -
    V_m \right] \Phi_m = -J_m = -a^{3/2} J  \cr
  \Phi_m = a^{3/2} \phi \qquad
  V_m(r) = m^2 + {k^2 \over a^2} + {3 \over 2} {a' \over a} f' +
    {3 \over 4} {a'^2 \over a^2} f + {3 \over 2} {a'' \over a} f \,,
 }
where primes denote $d/dr$ and $\square_2$ is the laplacian on the orbit spacetime
 \eqn{OrbitSpacetime}{
  ds_2^2 = -f dt^2 + {1 \over f} dr^2 \,.
 }
The first aim of this paper is to reduce the linearized Einstein equations, sourced by the trailing string, to five master equations, each one similar to \eno{MassiveMaster}.  The homogenous parts of these equations are well known, having been worked out in full in \cite{Kodama:2003jz} following earlier work including \cite{Bardeen:1980kt,Kodama:1985bj}.  The source terms were partly worked out in \cite{Kodama:2000fa}, whose methods we follow.\footnote{Our methods are also reminiscent of those used in \cite{Benincasa:2005iv} to study sound waves in ${\cal N}=2^*$ gauge theory via a gauge-invariant formalism.}  With an eye toward future applications to non-conformal backgrounds, we derive the master equations in the general warped background \eno{CurvedBackground} dual to a finite-temperature field theory on ${\bf R}^{3,1}$, without making assumptions about the stress tensor that supports the five-dimensional curvature.

With the master equations in hand, we specialize to the case of the trailing string in $AdS_5$-Schwarzschild and reproduce some of the results of \cite{Friess:2006fk}.  Also we consider an analytic approximation for the large $k$ region, following methods developed in \cite{Yarom:2007ap}.  We explain in appendix~\ref{JULYVARIABLES} how the parameterization of metric perturbations used in this paper relates to the axial gauge parameterization used in \cite{Friess:2006fk}.

The current work has some overlap with the independent study \cite{YaromNew}.

\section{Decoupling the linearized Einstein equations}
\label{DECOUPLING}

The first step in formulating the master equations is to expand the metric and stress tensor perturbations into Fourier harmonics involving scalar-, vector-, and tensor-valued functions on ${\bf R}^3$.  As we describe in section~\ref{HARMONICS}, the scalar case involves nothing but plane waves, and the other two cases involve plane waves times appropriate polarization tensors.  The coefficients of these harmonics are functions of $t$ and $r$ which in general have some tensor structure in the orbit space~\eno{OrbitSpacetime}.  We refer to terms in the Fourier expansions as scalar, vector, and tensor perturbations according to the type of harmonic involved rather than their tensor structure in the orbit space: indeed, one finds that the scalar perturbations are expressed in terms of a symmetric tensor in the orbit space; the vector perturbations can be expressed in terms of a vector in the orbit space; and the tensor perturbations can be expressed in terms of an orbit-space scalar.  In the case of tensor perturbations, the orbit-space scalar is (up to an overall factor) the master field $\Phi_T$, as described in section~\ref{TENSORMASTER}.  In the case of vector perturbations, the orbit-space vector can be expressed in terms of an orbit-space scalar $\Phi_V$ by use of one of the linearized Einstein equations, as described in section~\ref{VECTORMASTER}.  Likewise, in the case of scalar perturbations, the orbit-space tensor can be expressed in terms of an orbit-space scalar $\Phi_S$, as described in section~\ref{SCALARMASTER}.  Each master field satisfies a second order master equation, which implies the linearized Einstein equations not already used in constructing the master field.\footnote{We do not know of a general theorem that master fields satisfying second-order master equations can always be found.}  There is an intrinsic parity for tensor and vector perturbations which affects the polarization tensors but not the form of the master equation.  Thus the five master fields are $\Phi_T^{\rm even}$, $\Phi_T^{\rm odd}$,$\Phi_V^{\rm even}$, $\Phi_V^{\rm odd}$, and $\Phi_S$.  The corresponding master equations (without reference to parity) are given in \eno{MasterEQ},~\eno{MasterV}, and~\eno{MasterS}.

\subsection{Scalar, vector, and tensor harmonics}
\label{HARMONICS}

The defining equations for scalar, vector, and tensor harmonics on ${\bf R}^3$ are\\[-10pt]
 \eqn{ScalarDef}{
  (\partial_i \partial^i + k^2) \mathbb{S} = 0
 }
 \eqn{VectorDef}{
  (\partial_i \partial^i + k^2) \mathbb{V}_j = 0 \qquad  \partial^j \mathbb{V}_j = 0
 }
 \eqn{TensorDef}{
  (\partial_i \partial^i + k^2) \mathbb{T}_{jh} = 0 \qquad \partial^j \mathbb{T}_{jh} = 0 = \mathbb{T}^{j}_{\phantom{j}j}\,,
 }
where indices are raised and lowered using the standard metric on ${\bf R}^3$.

The scalar harmonics can be chosen to be pure exponentials:
 \eqn{GetScalar}{
  \mathbb{S}(\vec{k},\vec{x}) = e^{i\vec{k} \cdot \vec{x}} \,.
 }
They satisfy the normalization condition
 \eqn{NormScalar}{
  \langle \mathbb{S}(\vec{k}), \mathbb{S}(\vec{k'}) \rangle = (2\pi)^3 \delta^3(\vec{k} - \vec{k}') \qquad \langle \mathbb{S}_1, \mathbb{S}_2 \rangle \equiv \int_{\mathbf{R}^3} d^3 x\, \mathbb{S}_1^{*}(\vec{x})\, \mathbb{S}_2(\vec{x}) \,.
 }

Vector harmonics may be assumed to be proportional to $e^{i\vec{k} \cdot \vec{x}}$, but for each choice of wave-number $\vec{k}$ there are two distinct solutions, which are conventionally specified by their transformation property under the parity transformation $\vec{x} \to -\vec{x}$, $\vec{k} \to -\vec{k}$, which preserves both the defining equation \eno{VectorDef} and the exponential part of the harmonic.  Denoting $k = |\vec{k}|$ and $k_\perp = \sqrt{k_2^2+k_3^2}$, one may express the solutions as
 \eqn{GetVector}{
  \mathbb{V}^{\rm even}_i(\vec{k},\vec{x}) &= {1\over k k_\perp} e^{i \vec{k} \cdot \vec{x}} \begin{pmatrix} k_\perp^2 & -k_1 k_2 & -k_1 k_3 \end{pmatrix}\cr
  \mathbb{V}^{{\rm odd}}_i (\vec{k},\vec{x}) &= {1\over k_\perp} e^{i \vec{k} \cdot \vec{x}} \begin{pmatrix} 0 & -k_3 & k_2 \end{pmatrix}\,.
 }
These vector harmonics \eqref{GetVector} satisfy the normalization conditions
 \eqn[c]{NormVector}{
  \langle \mathbb{V}^{{\rm even}}(\vec{k}), \mathbb{V}^{{\rm even}}(\vec{k}') \rangle = (2\pi)^3 \delta^3(\vec{k} - \vec{k}') = \langle \mathbb{V}^{{\rm odd}}(\vec{k}), \mathbb{V}^{{\rm odd}}(\vec{k}') \rangle\cr
  \langle \mathbb{V}^{{\rm even}}(\vec{k}), \mathbb{V}^{{\rm odd}}(\vec{k}') \rangle = 0
 }
under the inner product
 \eqn{VectorInnerProd}{
  \langle \mathbb{V}_1, \mathbb{V}_2 \rangle = \int_{\mathbf{R}^3} d^3 x\, \mathbb{V}^*_{1, j}(\vec{x})\, \mathbb{V}_2^j(\vec{x}) \,.
 }
It is convenient for us to make the $x^1$ direction privileged, because our eventual aim is to describe the emission from an external quark moving in the $x^1$ direction through a thermal plasma of ${\cal N}=4$ gauge theory.  A general vector field $X_i(\vec{x})$ on ${\bf R}^3$ admits the Fourier expansion
 \eqn{GeneralVector}{
  X_i(\vec{x}) = \int {d^3 k \over (2\pi)^3} \left[
    X_V^{\rm even}(\vec{k})
      \mathbb{V}^{\rm even}_i(\vec{k},\vec{x}) +
    X_V^{\rm even}(\vec{k})
      \mathbb{V}^{\rm odd}_i(\vec{k},\vec{x}) +
    X_S(\vec{k}) \mathbb{S}_i(\vec{k},\vec{x}) \right] \,,
 }
for some set of Fourier coefficients $X_V^{\rm even}(\vec{k})$, $X_V^{\rm odd}(\vec{k})$, and $X_S(\vec{k})$, where
 \eqn{VectorMoreDef}{
  \mathbb{S}_i(\vec{k},\vec{x}) = -{1\over k} \partial_i \mathbb{S}(\vec{k},\vec{x})\,.
 }
The expansion \eno{GeneralVector} is orthonormal with respect to the inner product \eno{VectorInnerProd}.

Tensor harmonics may also be assumed to be proportional to $e^{i\vec{k} \cdot \vec{x}}$, and there are again two distinct polarization tensors with definite parity:
 \eqn{GetTensorEven}{
  \mathbb{T}^{{\rm even}}_{ij} (\vec{k},\vec{x}) = {1\over k^2} e^{i \vec{k} \cdot \vec{x}} \begin{pmatrix} k_\perp^2 & -k_1 k_2 & -k_1 k_3 \\[3\jot]
  -k_1 k_2 & -\displaystyle{-k_1^2 k_2^2 + k^2 k_3^2 \over k_\perp^2} & k_2 k_3 \displaystyle{2 k_1^2 + k_\perp^2 \over k_\perp^2}\\[5\jot]
  -k_1 k_3 & k_2 k_3 \displaystyle{2 k_1^2 + k_\perp^2 \over k_\perp^2} & -\displaystyle{-k_1^2 k_3^2 + k^2 k_2^2 \over k_\perp^2}
  \end{pmatrix}
 }
 \eqn{GetTensorOdd}{
  \mathbb{T}^{{\rm odd}}_{ij} (\vec{k},\vec{x}) = {1\over k} e^{i \vec{k} \cdot \vec{x}} \begin{pmatrix} 0 & -k_3 & k_2 \\[3\jot]
  -k_3 & \displaystyle{2 k_1 k_2 k_3\over k_\perp^2} & \displaystyle{k_1 (k_3^2 - k_2^2) \over k_\perp^2}\\[5\jot]
   k_2& \displaystyle{k_1 (k_3^2 - k_2^2) \over k_\perp^2} & -\displaystyle{2 k_1 k_2 k_3 \over k_\perp^2}
  \end{pmatrix}\,.
 }
With the inner product
 \eqn{TensorInnerProd}{
  \langle \mathbb{T}_1, \mathbb{T}_2 \rangle = {1 \over 2} \int_{\mathbf{R}^3} d^3 x\, \mathbb{T}^*_{1, jh}(\vec{x})\, \mathbb{T}_2^{jh}(\vec{x})
 }
we have the normalization condition
 \eqn[c]{NormTensor}{
  \langle \mathbb{T}^{{\rm even}}(\vec{k}), \mathbb{T}^{{\rm even}}(\vec{k'}) \rangle = (2\pi)^3 \delta^3(\vec{k} - \vec{k}') = \langle \mathbb{T}^{{\rm odd}}(\vec{k}), \mathbb{T}^{{\rm odd}}(\vec{k'}) \rangle\cr
  \langle \mathbb{T}^{{\rm even}}(\vec{k}), \mathbb{T}^{{\rm odd}}(\vec{k'}) \rangle = 0\,.
 }
A general symmetric tensor field $X_{ij}(\vec{x})$ on ${\bf R}^3$ admits the Fourier decomposition
 \eqn{Tdecompose}{
  X_{ij}(\vec{x}) &= \int {d^3 k \over (2\pi)^3} \bigg[
   X_L^S(\vec{k}) \delta_{ij} \mathbb{S}(\vec{k},\vec{x}) +
   X_T^S(\vec{k}) \mathbb{S}_{ij}(\vec{k},\vec{x})
       \cr &\qquad\qquad{} +
   X_V^{\rm even}(\vec{k})
     \mathbb{V}^{\rm even}_{ij}(\vec{k},\vec{x}) +
   X_V^{\rm odd}(\vec{k})
     \mathbb{V}^{\rm even}_{ij}(\vec{k},\vec{x})
       \cr &\qquad\qquad{} +
   X_T^{\rm even}(\vec{k})
     \mathbb{T}^{\rm even}_{ij}(\vec{k},\vec{x}) +
   X_T^{\rm odd}(\vec{k})
     \mathbb{T}^{\rm odd}_{ij}(\vec{k},\vec{x}) \bigg] \,,
 }
where
 \eqn[c]{TensorMoreDef}{
  \mathbb{V}^{{\rm even}}_{ij}(\vec{k},\vec{x}) = -{1\over k} \partial_{(i} \mathbb{V}^{{\rm even}}_{j)} \qquad
  \mathbb{V}^{{\rm odd}}_{ij}(\vec{k},\vec{x}) = -{1\over k} \partial_{(i} \mathbb{V}^{{\rm odd}}_{j)}  \cr
  \mathbb{S}_{ij}(\vec{k},\vec{x}) = {1\over k^2} \partial_{i} \partial_j \mathbb{S} + {1\over 3} \delta_{ij} \mathbb{S}\,,
 }
and we use the notation $(ij) = {1 \over 2} (ij + ji)$.  The decomposition \eno{Tdecompose} is orthogonal with respect to the inner product \eno{NormTensor}, but not orthonormal because of some $k$-independent factors arising from the inner products of the derived harmonics in \eno{TensorMoreDef}.

\subsection{Fourier decomposition of the perturbations}
\label{DECOMPOSE}

The background \eno{CurvedBackground} is, by assumption, a solution of the five-dimensional Einstein equations,
 \eqn{EinsteinEqs}{
  R_{\mu\nu} - {1 \over 2} R G_{\mu\nu} + \Lambda G_{\mu\nu} =
    T_{\mu\nu} \,.
 }
The explicit cosmological term is redundant because it could have been soaked into $T_{\mu\nu}$; however, retaining it explicitly makes it easier to apply the formalism to an anti-de Sitter space example.  $T_{\mu\nu}$ includes any contributions from bulk scalar fields or other matter.  Let the background metric and stress tensor be denoted $G^{(0)}_{\mu\nu}$ and $T^{(0)}_{\mu\nu}$.  Consider a perturbation of both the metric and the stress tensor:
 \eqn{GmunuSeries}{
  G_{\mu\nu} = G^{(0)}_{\mu\nu} + \lambda h_{\mu\nu} \qquad
  T_{\mu\nu} = T^{(0)}_{\mu\nu} + \lambda \tau_{\mu\nu} \,,
 }
where $\lambda$ is a formal expansion parameter which we eventually set to $1$.

Let indices $b$ and $c$ run over $t$ and $r$, while indices $i$ and $j$ run over the three $\vec{x}$ directions.  The metric perturbations can be decomposed into three functions $h_{bc}$ which are scalars on ${\bf R}^3$, two vector-valued functions $h_{bi}$, and one tensor-valued function $h_{ij}$.  These in turn may be expanded in Fourier series as indicated in \eno{GeneralVector} and \eno{Tdecompose}:
 \eqn{hbcDecomp}{
  h_{bc}(t,r,\vec{x}) = \int {d^3 k\over (2 \pi)^3} f_{bc}^S(\vec{k},t,r)\, \mathbb{S}(\vec{k},\vec{x})
 }
 \eqn{hbiDecomp}{
  h_{bi}(t,r,\vec{x}) &= a(r) \int {d^3 k\over (2 \pi)^3} \Big[f_b^S(\vec{k},t,r)\, \mathbb{S}_i(\vec{k},\vec{x}) \cr&\qquad\qquad{} + f_b^{V, {\rm even}}(\vec{k},t,r)\, \mathbb{V}_i^{{\rm even}}(\vec{k},\vec{x}) + f_b^{V, {\rm odd}}(\vec{k},t,r)\, \mathbb{V}_i^{{\rm odd}}(\vec{k},\vec{x})\Big]
 }
 \eqn{hijDecomp}{
  h_{ij}(t,r,\vec{x}) = 2 a(r)^2 \int &{d^3 k\over (2 \pi)^3} \bigg[ H_L^S(\vec{k},t,r) \delta_{ij}\, \mathbb{S}(\vec{k},\vec{x}) + H_T^S(\vec{k},t,r)\, \mathbb{S}_{ij}(\vec{k},\vec{x}) \cr&\qquad{} + H_T^{V, {\rm even}}(\vec{k},t,r)\, \mathbb{V}_{ij}^{{\rm even}}(\vec{k},\vec{x}) + H_T^{V, {\rm odd}}(\vec{k},t,r)\, \mathbb{V}_{ij}^{{\rm odd}}(\vec{k},\vec{x}) \cr&\qquad{} + H_T^{T, {\rm even}}(\vec{k},t,r)\, \mathbb{T}_{ij}^{{\rm even}}(\vec{k},\vec{x}) + H_T^{T, {\rm odd}}(\vec{k},t,r)\, \mathbb{T}_{ij}^{{\rm odd}}(\vec{k},\vec{x})\bigg]\,,
 }
where the factors of $a(r)$ are chosen for convenience.  Likewise one may expand
 \eqn{taubcDecomp}{
  \tau_{bc}(t,r,\vec{x}) = \int {d^3 k\over (2 \pi)^3} \tau_{bc}^S(\vec{k}) \mathbb{S}(\vec{k},\vec{x})
 }
 \eqn{taubiDecomp}{
 \tau_{bi}(t,r,\vec{x}) &= a(r) \int {d^3 k\over (2 \pi)^3} \bigg[\tau_b^S(\vec{k},t,r) \mathbb{S}_i(\vec{k},\vec{x}) \cr&\qquad\qquad{} + \tau_b^{V, {\rm even}}(\vec{k},t,r)  \mathbb{V}_i^{{\rm even}}(\vec{k},\vec{x}) + \tau_b^{V, {\rm odd}}(\vec{k},t,r) \mathbb{V}_i^{{\rm odd}}(\vec{k},\vec{x})\bigg]
 }
 \eqn{tauijDecomp}{
  \tau_{ij}(t,r,\vec{x}) &= 2 a(r)^2 \int {d^3 k\over (2 \pi)^3} \bigg[ p^S(\vec{k},t,r) \delta_{ij}\, \mathbb{S}(\vec{k},\vec{x}) + \tau^S(\vec{k},t,r)\, \mathbb{S}_{ij}(\vec{k},\vec{x}) \cr&\qquad\qquad{} + \tau^{V, {\rm even}}(\vec{k},t,r)\, \mathbb{V}_{ij}^{{\rm even}}(\vec{k},\vec{x}) + \tau^{V, {\rm odd}}(\vec{k},t,r)\, \mathbb{V}_{ij}^{{\rm odd}}(\vec{k},\vec{x}) \cr&\qquad\qquad{} + \tau^{T, {\rm even}}(\vec{k},t,r)\, \mathbb{T}_{ij}^{{\rm even}}(\vec{k},\vec{x}) + \tau^{T, {\rm odd}}(\vec{k})\, \mathbb{T}_{ij}^{\vec{k}, {\rm odd}}(\vec{x})\bigg]\,.
 }
It is also useful to express a general vector $v_\mu$ as
 \eqn{vParam}{
  v_b &= \int {d^3 k \over (2\pi)^3}
    v_b^S(\vec{k},t,r) \mathbb{S}(\vec{k},\vec{x})  \cr
  v_i &= a(r)^2 \int {d^3 k \over (2\pi)^3}
    \bigg[ v^{V,\rm even}(\vec{k},t,r)
       \mathbb{V}^{\rm even}_i(\vec{k},\vec{x}) +
      v^{V,\rm odd}(\vec{k},t,r)
       \mathbb{V}^{\rm odd}_i(\vec{k},\vec{x})
     \cr &\qquad\qquad{} +
      v_V^S(\vec{k},t,r)
       \mathbb{S}_i(\vec{k},\vec{x}) \bigg] \,.
 }
The diffeomorphism symmetry of the Einstein equations \eno{EinsteinEqs} may be expressed as
 \eqn{Diffeomorphisms}{
  \delta_v G_{\mu\nu} &= \lambda {\cal L}_v G_{\mu\nu}
    = \lambda (\nabla_\mu v_\nu + \nabla_\nu v_\mu)  \cr
  \delta_v T_{\mu\nu} &= \lambda {\cal L}_v T_{\mu\nu}
    = \lambda (v^\rho \partial_\rho T_{\mu\nu} +
       T_{\rho\nu} \partial_\mu v^\rho +
       T_{\mu\rho} \partial_\nu v^\rho) \,,
 }
where $v^\mu$ is an arbitrary vector, and for convenience we have quantified the smallness of the coordinate deformation in terms of the same formal expansion parameter $\lambda$ that we used in the metric expansion \eno{GmunuSeries}.  Thus, at linearized level, \eno{Diffeomorphisms} becomes
 \eqn{LinDiff}{
  \delta_v h_{\mu\nu} = \nabla_\mu^{(0)} v_\nu + \nabla_\nu^{(0)}
    v_\mu \qquad
  \delta_v \tau_{\mu\nu} = v^\rho \partial_\rho T_{\mu\nu}^{(0)} +
    T_{\rho\nu}^{(0)} \partial_\mu v^\rho + T_{\mu\rho}^{(0)}
     \partial_\nu v^\rho \,.
 }
For the purpose of formulating master equations for metric perturbations, we do not need to know details about how the zeroth order stress tensor arises from matter fields.  Instead, we may extract $T_{\mu\nu}^{(0)}$ in terms of the unperturbed metric using the zeroth order Einstein equations:
 \eqn{Tzero}{
  T_{\mu\nu}^{(0)} &= R_{\mu\nu}^{(0)} -
    {1 \over 2} R^{(0)} G_{\mu\nu}^{(0)} +
    \Lambda G_{\mu\nu}^{(0)}
   = \diag\{ T_{tt}^{(0)}, T_{rr}^{(0)}, T_{xx}^{(0)}, T_{xx}^{(0)},
       T_{xx}^{(0)} \}  \cr
  T_{tt}^{(0)} &= -\Lambda f - 3 {a'^2 \over a^2} f^2 -
    {3 \over 2} {a' \over a} ff' - 3 {a'' \over a} f^2  \cr
  T_{rr}^{(0)} &= {\Lambda \over f} + 3 {a'^2 \over a^2} +
    {3 \over 2} {a' \over a} {f' \over f}  \cr
  T_{xx}^{(0)} &= \Lambda a^2 + fa'^2 + 2aa'f' + 2aa''f + {1 \over 2}
    a^2 f'' \,.
 }

\subsection{Master equation for the tensor modes}
\label{TENSORMASTER}

Decoupled equations of motion for the tensor modes may be found directly by plugging the expansions \eno{hijDecomp} and \eno{tauijDecomp} into the linearized Einstein equations.  The result is the same for even and odd modes, so we will simply omit to specify parity in the following.  The master equation is
 \eqn{MasterEQ}{
  \left( \square_2 - V_T \right) \Phi_T = -J_T
 }
where
 \eqn[c]{MasterFieldT}{
  \Phi_T = a^{3/2} H_T^T \qquad
   J_T = 2 a^{3/2} \tau^T  \cr
  V_T = 2\Lambda + {k^2 \over a^2} + {11 \over 2} {a' \over a} f' +
    {11 \over 4} {a'^2 \over a^2} f +
    {11 \over 2} {a'' \over a} f
 }
and, as before, $\square_2 = -{1\over f} \partial_t^2 + \partial_r f \partial_r$ is the laplacian on the two-dimensional orbit spacetime \eno{OrbitSpacetime}.

\subsection{Master equation for the vector modes}
\label{VECTORMASTER}

The results for vector modes are the same for even and odd parity, so we will not refer to parity explicitly in the rest of this section.  As explained in \cite{Kodama:2000fa,Kodama:2003jz}, the first step is to consider quantities that are invariant under diffeomorphisms.  To this end, consider the transformation properties of vector components of the metric and tensor perturbations, as can be derived by substituting the expansions \eno{hbcDecomp}--\eno{vParam} into \eno{LinDiff}:
 \eqn{DiffAction}{\seqalign{\span\TL & \span\TR &\qquad\span\TL & \span\TR}{
  f_b^V &\to f_b^V + a \partial_b v_V &
   H_T^V &\to H_T^V - k v_V  \cr
  \tau_b^V &\to \tau_b^V + {T_{xx}^{(0)} \over a} \partial_b v_V &
   \tau^V &\to \tau^V - {k T_{xx}^{(0)} \over a^2} v_V \,.
 }}
One may form diffeomorphism-invariant combinations as follows:
 \eqn{FaDef}{
  \hat{f}_b^V = f_b^V + {a \over k} \partial_b H_T^V \qquad
  \hat\tau_b^V = \tau_b^V + {T_{xx}^{(0)} \over ka} \partial_b H_T^V \qquad
  \hat\tau^V = \tau^V - {T_{xx}^{(0)} \over a^2} H_T^V \,.
 }
If the $\hat{f}_b^V$ are expressed in terms of the master field as follows:\footnote{Two degrees of freedom, namely $\hat{f}_t^V$ and $\hat{f}_r^V$, can be reduced to one, namely $\Phi_V$, because of a first order constraint that would emerge if we substituted general $\hat{f}_b^V$ into the linearized Einstein equations.  This is explained, for example, in section IV.B.2 of \cite{Kodama:2000fa}.}
 \eqn{FaVSmaster}{
  \partial_t \hat{f}^t_V = -{2a \over k} \hat\tau^V -
    {1 \over a^2} \partial_r (a^{3/2} \Phi_V) \qquad
  \hat{f}^r_V = {\Phi_V \over \sqrt{a}} \,,
 }
then the $bc$ and $ij$ components of the linearized Einstein equations are satisfied automatically.  The $tj$ components lead to an equation for $\Phi_V$ involving up to third order derivatives in $r$, while the $rj$ components lead to a second order equation for $\Phi_V$.  In deriving these equations, we found it convenient to pass to a gauge where $H_T^V=0$, and to assume harmonic time dependence, $e^{-i\omega t}$, for $\hat{f}_b^V$, $\hat\tau_b^V$, $\hat\tau^V$, and $\Phi_V$ (which are now indistinguishable from the corresponding unhatted quantities).  A crucial point is that the third order equation for $\Phi_V$ follows from the second order equation plus a relation that follows from the conservation of the stress tensor.  More explicitly, from the ${\cal O}(\lambda)$ term in the equation
 \eqn{ConserveT}{
  \left( \nabla^\mu_{(0)} + \lambda \nabla^\mu_{(1)} \right)
    \left( T_{\mu\nu}^{(0)} + \lambda \tau_{\mu\nu} \right) = 0
 }
one may derive the relation in question:
 \eqn{ConserveTauV}{
  {1 \over a^4} D^b (a^4 \hat\tau^V_b) &=
    -{k^2 + 2 T_{xx}^{(0)} \over ka} \hat\tau^V +
    {\partial_r T_{xx}^{(0)} \over a^{5/2}} \Phi_V
 }
where $D_b$ is the covariant derivative with respect to the orbit spacetime \eno{OrbitSpacetime}.

The master equation for the vector modes is just the second order equation following from the $rj$ Einstein equations:
 \eqn{MasterV}{
  \left( \square_2 - V_V \right) \Phi_V = -J_V
 }
where
 \eqn{MasterSourceV}{
  V_V &= 2\Lambda + {k^2 \over a^2} + {5 \over 2} {a' \over a} f' +
    {23 \over 4} {a'^2 \over a^2} f + {5 \over 2} {a'' \over a} f +
    f''  \cr\noalign{\vskip2\jot}
  J_V &= 2 \sqrt{a} f \left[ \hat\tau_r^V +
    {a \over kf} \partial_r (f \hat\tau^V) \right] \,.
 }

\subsection{Master equation for the scalar modes}
\label{SCALARMASTER}

For scalar modes, there is no longer a notion of parity.  The diffeomorphism transformations of the metric perturbations are
 \eqn{DiffScalar}{\seqalign{\span\TL & \span\TR & \qquad\span\TL & \span\TR}{
  f_{bc}^S &\to f_{bc}^S + 2 D_{(b} v_{c)}^S &
  f_b^S &\to f_b^S - {k \over a} v_b^S + a \partial_b v_V^S  \cr
  H_T^S &\to H_T^S - k v_V^S &
  H_L^S &\to H_L^S + {k \over 3} v_V^S +
    {a' \over a} f v_r^S \,.
 }}
Because of the appearance of $v_b^S$ without derivatives in the transformation of $f_b^S$ and of $v_V^S$ without derivatives in the transformation of $H_T^S$, these quantities may be gauged away or used to construct diffeomorphism-invariant variants of $f_{bc}^S$ and $H_L^S$:
 \eqn{ScalarInvariants}{
  \hat{f}_{bc}^S = f_{bc}^S + 2 D_{(b} X_{c)} \qquad
  \hat{H}_L^S = H_L^S + {1 \over 3} H_T^S +
    {a' \over a} f X_r
 }
where
 \eqn{XaDef}{
  X_b = {a \over k} \left( f_b^S + {a \over k} \partial_b
    H_T^S \right)
 }
is invariant under diffeomorphisms with only $v_V^S$ non-zero.  Also,
 \eqn{MoreDiffScalar}{
  \tau_{bc}^S &\to \tau_{bc}^S + v^{S, d} \partial_d T_{bc}^{(0)} + 2 T_{d(b}^{(0)} \partial_{c)} v^{S, d}\cr
  \tau_{b}^S &\to \tau_b^S - {k \over a} T_{bd}^{(0)} v^{S, d} + {1\over a} T_{xx}^{(0)} \partial_b v_V^S\cr
  \tau^S &\to \tau^S - {k\over a^2} T_{xx}^{(0)} v_V^S\cr
  p^S &\to p^S + {k \over 3 a^2} T_{xx}^{(0)} v_V^S + {f \over 2 a^2} v_r^S \partial_r T_{xx}^{(0)}
 }
and the gauge-invariant quantities are
 \eqn{MoreScalarInvariants}{
  \hat{\tau}_{bc}^S &= \tau_{bc}^S + X^{d} \partial_d T_{bc}^{(0)} + 2 T_{d(b}^{(0)} \partial_{c)} X^d\cr
  \hat{\tau}_{b}^S &= \tau_b^S - {k \over a} T_{bd}^{(0)} X^{d} + {1\over k a} T_{xx}^{(0)} \partial_b H_T^S\cr
  \hat{\tau}^S &= \tau^S - {1\over a^2} T_{xx}^{(0)} H_T^S\cr
  \hat{p}^S &= p^S + {1 \over 3 a^2} T_{xx}^{(0)} H_T^S + {f \over 2 a^2} X_r^S \partial_r T_{xx}^{(0)} \,.
 }

In manipulating the linearized Einstein equations, we employed a gauge where $f_b^S = 0$ and $H_T^S=0$.  One may however pass immediately to diffeomorphism-invariant formulas by replacing $f_{bc}^S$, $H_L^S$, $\tau^S$, $\tau^S_b$, and $\tau^S_{bc}$ by their diffeomorphism-invariant relatives as defined in \eno{ScalarInvariants} and \eno{MoreScalarInvariants}, and we will quote all formulas in diffeomorphism-invariant form.  One may eliminate $\hat{H}_L^S$ by noting that the $ij$ Einstein equations with $i \neq j$---that is, the off-diagonal equations in the lower-right $3 \times 3$ block---lead directly to
 \eqn{HLSelim}{
  \hat{H}_L^S = -{1 \over 2} G_{(0)}^{bc} \hat{f}^S_{bc} -
    {2a^2 \over k^2} \hat\tau^S \,.
 }
The $tr$, $rr$, $ti$, and $ri$ Einstein equations involve only first order derivatives of $\hat{f}_{bc}^S$, and using stress tensor conservation equations one may show that the other Einstein equations are implied by these four.  Approximately following \cite{Kodama:2003jz} we express the symmetric tensor $\hat{f}_{bc}^S$ as
 \eqn{ExpressFabS}{
  \hat{f}_{tt}^S = {f \over 3a} (2X-Y) \qquad
  \hat{f}_{tr}^S = {i\omega \over af} Z \qquad
  \hat{f}_{rr}^S = {1 \over 3af} (-2Y+X) \,.
 }
Then the first order equations just mentioned reduce to
 \eqn[c]{FullXYZ}{
  X' - {a' \over a} X + \left( 2 {a' \over a} - {f' \over f}
    \right) Y + {1 \over f} \left( -{\omega^2 \over f} +
     {k^2 \over a^2} - {2 T_{tt}^{(0)} \over f}
       \right) Z = \tau_X  \cr\noalign{\vskip2\jot}
  Y' - {f' \over 2f} X + {f' \over 2f} Y +
    {\omega^2 \over f^2} Z = \tau_Y \qquad\qquad
  Z' + X = \tau_Z  \cr\noalign{\vskip2\jot}
  \left( \omega^2 + {f'^2 \over 4} +
    {2\Lambda \over 3} f + {3 \over 2} {a' \over a} ff' \right) X +
  \left( \omega^2 + {f'^2 \over 4} - {k^2 \over a^2} f -
    {4\Lambda \over 3} f - 3 {a'^2 \over a^2} f^2 -
    3 {a' \over a} ff' \right) Y  \cr{} +
  \left( 3\omega^2 {a' \over a} - {k^2 \over 2} {f' \over a^2} +
    T_{tt}^{(0)} {f' \over f} \right) Z = \tau_C
 }
where
 \eqn{tauXYZ}{
  \tau_X &= {2 \over i\omega} a \hat\tau^S_{tr} -
    {2a^2 \over k} \hat\tau^S_r +
    {2a^2 \over k^2} \left( 2a \partial_r -
      3a {f' \over f} + 10a' \right) \hat\tau^S  \cr
  \tau_Y &= {2a^2 \over k} \hat\tau^S_r +
    {8a^2 \over k^2} \left( a \partial_r + 2 a' \right)
     \hat\tau^S  \cr
  \tau_Z &= -{2a^2 \over ik\omega} \hat\tau^S_t +
    {8a^3 \over k^2} \hat\tau^S  \cr
  \tau_C &= -{aff' \over i\omega} \hat\tau^S_{tr} +
    2af^2 \hat\tau^S_{rr} - {6aa'f^2 \over k} \hat\tau^S_r +
    {a^3 \over k^2} \left( 12\omega^2 + 3 f'^2 - {8k^2 \over a^2} f -
     {6fa'f' \over a} \right) \hat\tau^S \,.
 }
Because the system of equations \eqref{FullXYZ} consists of three first order differential equations and one algebraic constraint, it can be reduced to a single second order differential equation satisfied by any desired linear combination of $X$, $Y$, and $Z$.  In order to put this equation in a form that resembles the other master equations, we define the scalar master field as
 \eqn{MasterSDef}{
  \Phi_S = A \left( X + Y + 3 {a'\over a} Z \right)\,,
 }
where
 \eqn{ADef}{
  A \equiv {a^{1/2}\over H^{1/2} (H + 3 a f a'')^{1/2}} \qquad H \equiv k^2 + 2 \Lambda a^2 + 3 f a'^2 + {9\over 2} a' f'\,.
 }
The master equation then reads:
 \eqn{MasterS}{
  \left( \square_2 + W_S \partial_r - V_S \right) \Phi_S = -J_S\,,
 }
with $\square_2 = {1\over f} \omega^2 + \partial_r f \partial_r$ and
 \eqn{WSDef}{
  W_S = {3 f\over 2 H}\left(4 \Lambda a a' + 9 f' a'^2 + 4 f a' a'' + 5 a f' a'' + 3 a a' f'' \right)\,.
 }
In the general warped background \eqref{CurvedBackground}, the expressions for $V_S$ and $J_S$ are too long to be reproduced here.  Starting in section~\ref{TRAILING}, we specialize to the case where the unperturbed metric \eqref{CurvedBackground} is $AdS_5$-Schwarzschild, namely $a = r/L$ and $h = 1- r_H^4/r^4$, where $\Lambda = -6/L^2$.  In this case $W_S = 0$, so the master equation \eqref{MasterS} takes the standard form $(\square_2 - V_S) \Phi_S = -J_S$ where
 \eqn{ScalarDefs}{
  V_S &= {1\over 4 L^{10} r^8 H^2} \bigg[12 k^2 L^4 r^2 r_H^4 (9r^4 - 13 r_H^4) + k^4 L^8 r^4 (r^4-9 r_H^4) - 4 k^6 L^{12} r^6\cr
  &\hspace{1in}- 108 r_H^8 (5 r^4 + 3 r_H^4) \bigg] \cr
  J_S &= {f\over a^{3/2} H}\bigg[{a^{2}\over f^2} \tau_C + \left({24 r_H^4\over H L^6 r} + {2 r^4 + 4 r_H^4\over f L^4 r} \right)\left(\tau_X + \tau_Y\right) + a^2 \left(\tau_X' + \tau_Y' + {3\over r} \tau_Z' \right)\cr
  &\hspace{1in}+\left({72 r_H^4\over H L^6 r^2} + {3(r^4 + 5 r_H^4)\over L^4 r^2}-{H\over f}\right)\tau_Z \bigg]\cr
  H &= k^2 + {6 r_H^4\over L^4 r^2}\,.
 }
(Note that $H$ as quoted in \eno{ScalarDefs} is the same one defined in \eno{ADef}.)

\section{The trailing string in $AdS_5$-Schwarzschild}
\label{TRAILING}

Now, as promised, we turn to an application of the master field formalism to computations of $\langle T_{mn} \rangle$ as sourced by the trailing string configuration in $AdS_5$-Schwarzschild.  In section~\ref{TRAILING_VALUES} we work out the stress tensor in five dimensions created by the trailing string, in the Fourier-expanded forms \eno{taubcDecomp}--\eno{tauijDecomp}.  The resulting expressions may be inserted into the master equations \eno{MasterEQ}, \eno{MasterV}, and~\eno{MasterS} to obtain explicit differential equations in $r$ for the radial dependence of the master fields, which we give explicitly in \eno{psiIEQ}--\eno{JSTldDef}.  In sections~\ref{TENSOR_LIMITS}--\ref{SCALAR_LIMITS} we obtain asymptotic forms near the horizon and near the boundary of $AdS_5$-Schwarzschild for the solutions of the master equations.  As explained in section~\ref{HOLOGRAPHIC_STRESS}, the asymptotic forms near the boundary translate into the desired quantities $\langle T_{mn} \rangle$ in the gauge theory.

\subsection{The stress tensor of the trailing string}
\label{TRAILING_VALUES}

In Poincar\'e coordinates, the $AdS_5$-Schwarzschild metric can be realized as a particular case of \eqref{CurvedBackground} with
 \eqn{AdS5Schwarzschild}{
  a(r) = {r\over L} \qquad h(r) = 1 - {r_H^4 \over r^4}\,.
 }
Here, $r_H$ denotes the location of the black hole horizon, and the AdS radius $L$ is related to the cosmological constant term via
 \eqn{LAndLambda}{
  \Lambda = -{6\over L^2}\,.
 }
It is useful to note that horizon temperature is $T = r_H/\pi L^2$.  The equations of motion follow from the action
 \eqn{Action}{
  S = \int d^5 x \left[{\sqrt{-G} (R + 12/L^2) \over 2 \kappa_5^2} - {1\over 2 \pi \alpha'} \int d^2\sigma \sqrt{-\det g_{\alpha \beta}}\, \delta^5 \left(x^\mu - X^\mu(\sigma) \right)\right] \,,
 }
where $g_{\alpha\beta}$ is the worldsheet metric.  In static gauge, the worldsheet metric that minimizes the action is, to leading order, given by \cite{Herzog:2006gh,Gubser:2006bz}
 \eqn[c]{WorldsheetMetric}{
  g_{\alpha\beta} \equiv G_{\mu\nu} \partial_\alpha X^\mu \partial_\beta X^\nu \qquad X^\mu (t, r) \equiv \begin{pmatrix} t & r & X^1(t, r) & 0 & 0 \end{pmatrix}\cr
  X^1(t, r) = vt + \xi(r) \qquad \xi(r) = -{L^2 v\over 4 i r_H} \left(\log{r-i r_H\over r + i r_H} + i \log{r + r_H\over r - r_H} \right)\,.
 }
We can use the above expressions to compute the source term $T_{\mu\nu}$ that appears in the Einstein equations \eqref{EinsteinEqs}.  Because $AdS_5$-Schwarzschild is the solution to the Einstein equations with negative cosmological constant but vanishing stress-energy tensor of the matter fields, $T_{\mu\nu}$ comes only from the trailing string: that is,
 \eqn{TauDef}{
  T^{\mu\nu} = \tau^{\mu\nu} = {\kappa_5^2 \over 2 \pi \alpha'} {1\over \sqrt{1-v^2}} {L^5 \over r^5} \delta^3(x^i - X^i) \begin{pmatrix}
   \displaystyle{h + v^2 r_H^4/r^4\over h^2} & -\displaystyle{r_H^2 v^2 \over L^2 h} & \displaystyle{v\over h} & 0 & 0\\[4\jot]
   -\displaystyle{r_H^2 v^2\over L^2 h} & \displaystyle{r^4\over L^4}(v^2-h) & -\displaystyle{r_H^2 v\over L^2} & 0 & 0\\[3\jot]
   \displaystyle{v\over h} & -\displaystyle{r_H^2 v\over L^2} & v^2 & 0 & 0\\[3\jot]
   0 & 0 & 0 & 0 & 0\\
   0 & 0 & 0 & 0 & 0
   \end{pmatrix}\,,
 }
where we have set the formal expansion parameter $\lambda = 1$.  The notation $\delta^3(x^i - X^i)$ is short for $\delta\left(x^1 - v t - \xi(r) \right) \delta (x^2) \delta (x^3)$.  The expression \eno{TauDef} (with indices lowered) can then be decomposed as in \eqref{taubcDecomp}--\eqref{tauijDecomp}, with the coefficients of the decomposition given by
 \eqn{GotOddCoefficients}{
  \tau_{a}^{V, {\rm odd}} = \tau^{V, {\rm odd}} = \tau^{T, {\rm odd}} = 0
 }
 \eqn{GotijCoefficients}{
  p^S = {\cal T} {r^2 v^2 \over 6 L^2} \qquad \tau^S = {\cal T} {r^2 v^2 \left(k^2 - 3 k_1^2 \right)\over 4 k^2 L^2}\cr
  \tau^{V, {\rm even}} = {\cal T} {i  v^2 k_1 k_\perp r^2 \over k^2 L^2} \qquad \tau^{T, {\rm even}} = {\cal T} {v^2 k_\perp^2 r^2 \over 4 k^2 L^2}
 }
 \eqn{GotaiCoefficients}{
  \begin{pmatrix} \tau^S_t\\ \tau^S_r\end{pmatrix} = -{\cal T} {i v k_1 L\over k r h} \begin{pmatrix} h r^4/L^4\\r_H^2/L^2 \end{pmatrix} \qquad
  \begin{pmatrix} \tau^{V, {\rm even}}_t\\ \tau^{V, {\rm even}}_r\end{pmatrix} = -{\cal T} {v k_\perp L \over k r h} \begin{pmatrix} h r^4/L^4\\r_H^2/L^2 \end{pmatrix}
 }
 \eqn{GotabCoefficients}{
  \begin{pmatrix} \tau^S_{tt}  & \tau^S_{tr} \\ \tau^S_{rt} & \tau^S_{rr} \end{pmatrix} = {\cal T} {1\over h^2} \begin{pmatrix} h^2 (v^2 + h - v^2 h) r^4/L^4 & h v^2 r_H^2/L^2\\ h v^2 r_H^2/L^2 & v^2 - h \end{pmatrix}\,,
 }
where the common prefactor ${\cal T}$ is
 \eqn{CalTDef}{
  {\cal T} = \ell_v {L^5 \over r^5} e^{-i k_1 \left(v t + \xi(r)\right)}\qquad \ell_v \equiv {\kappa_5^2 \over 2 \pi \alpha'} {1\over \sqrt{1-v^2}}\,.
 }
The odd components are zero because of the choice of the even and odd vector and tensor harmonics that we made in section~\ref{HARMONICS}.  As a result, the odd-parity master equations may be solved trivially by setting $\Phi^{\rm odd}_T = \Phi^{\rm odd}_V = 0$.  We will henceforth be concerned only with the even parity cases.

The time-dependence of the source is $e^{-i v k_1 t}$, and we will make the steady-state assumption that the master fields $\Phi_T$, $\Phi_V$, and $\Phi_S$ have a similar time dependence.  It is convenient to make the following definitions:
 \eqn{RedefineMaster}{
  \Phi_T &= {\kappa_5^2 L^5\over 2} a^{3/2} \psi_T(r) e^{-i v k_1 t}
    \cr
  \Phi_V &= {2 i \kappa_5^2 v k_1 L^3\over k^2}  a^{3/2} \psi_V(r) e^{-i v k_1 t}
    \cr
  \Phi_S &= -{6 \kappa_5^2 L\over k^4} a^{3/2} \psi_S(r) e^{-i v k_1 t} \,.
 }
The equations satisfied by $\psi_I$ with $I= T$, $V$, and $S$ can be written as
 \eqn{psiIEQ}{
  \left[ {1 \over a^3} \partial_r a^3 f \partial_r +
    {v^2 k_1^2 \over f}  -
   {k^2 \over a^2} + {4-\nu_I^2 \over L^2} + \tilde{V}_I \right] \psi_I = -\tilde{J}_I \,,
 }
where $\nu_T = 2$, $\nu_V = 1$, and $\nu_S = 0$ (compare with equation \eqref{WaveEqnK}).  The quantities $\tilde{V}_I$ and $\tilde{J}_I$ are given by
 \eqn{VTldDefs}{
  \tilde{V}_T &= 0 \qquad \tilde{V}_V = {9 r_H^4\over L^2 r^4}\qquad \tilde{V}_S = {12 r_H^4\over H^2} \left[{2 k^2 \over L^6 r^2} + {k^4 \over L^2 r^4} - {12 r_H^4\over L^{10} r^4} + {2 k^2 r_H^4\over L^6 r^6}  \right]
 }
 \eqn{JTTldDef}{
  \tilde{J}_T &= {\ell_v\over \kappa_5^2} {v^2 k_\perp^2 e^{-i k_1 \xi}\over k^2 L^2 r^3}
 }
 \eqn{JVTldDef}{
  \tilde{J}_V = {\ell_v\over \kappa_5^2} {k_\perp e^{-i k_1 \xi}\over k k_1}  \left[-{v k_1\over L^2 r^2} + i r_H^2 (k^2 - v^2 k_1^2) {1\over r^5}  + {5 v k_1 r_H^4 \over L^2 r^6} \right]
 }
 \eqn{JSTldDef}{
  \tilde{J}_S &= {\ell_v\over \kappa_5^2} {e^{-i k_1 \xi} \over 6 k_1 L^6 H^2}  \bigg[k_1 \left(k^2 (2+ v^2) - 3 v^2 k_1^2\right) \left(-3 k^2 L^4 {1\over r} + \left(k^2 L^8 (2 k^2 - 3 v^2 k_1^2)-90 r_H^4\right) {1\over r^3}\right)\cr
  &\qquad{}- 3 i v k^2 L^6 r_H^2 \left(2 k^4 + v^2 k_1^2 (k^2 - 3 k_1^2) \right) {1\over r^4} + 3 k_1 L^4 r_H^4 \Big(18 v^4 k_1^4 + k^4 (2-5 v^2) \cr
  &\qquad{}+ 3  v^2 k^2 k_1^2 (1-2v^2) \Big) {1\over r^5} + 6 i v r_H^6 L^2 \left(2 k^4 + 9 v^2 k_1^2 (k^2 - 3 k_1^2) \right) {1\over r^6}\cr
  &\qquad{}+ 18 k_1 r_H^8 \left(k^2 (2 + 3 v^2) - 9 v^2 k_1^2 \right){1\over r^7}\bigg]\,,
 }
where $H = k^2 + 6r_H^4/L^4 r^2$ as in \eno{ScalarDefs}.  Note that the $\tilde{V}_I$ decay at the boundary as $1/r^2$ or faster.

\subsection{The holographic stress tensor}
\label{HOLOGRAPHIC_STRESS}

To understand the holographic relation of the master fields to the gauge theory stress tensor $T_{mn}$, a useful preliminary is to count the independent components of $T_{mn}$.  Because the stress tensor is symmetric, we start with $10$ components.  Conservation eliminates four, and conformal symmetry eliminates one more, so there are five left.  These five independent components are related holographically to the five master fields.  One can be more explicit by expanding
 \eqn{TabExpand}{
  \langle T_{00} \rangle_{\rm conserved} = \int {d^3 k \over (2\pi)^3}
    q_S(\vec{k},t) \mathbb{S}(\vec{k},\vec{x})
 }
 \eqn{TaiExpand}{
  \langle T_{0i} \rangle_{\rm conserved} &=
   \int {d^3 k \over (2\pi)^3}
    \bigg[ {1 \over k} \partial_t q_S(\vec{k},t) \mathbb{S}_i
      + q_V^{\rm even}(\vec{k},t)
      \mathbb{V}_i^{\rm even}(\vec{k},\vec{x}) +
     q_V^{\rm odd}(\vec{k},t)
      \mathbb{V}_i^{\rm odd}(\vec{k},\vec{x}) \bigg]
 }
 \eqn{TijExpand}{
  \langle T_{ij} \rangle_{\rm conserved} &=
   2 \int {d^3 k \over (2\pi)^3}
    \bigg[ {q_S(\vec{k},t) \over 6} \delta_{ij}
       \mathbb{S}(\vec{k},\vec{x}) +
       {1 \over 4k^2} (3\partial_t^2 + k^2) q_S(\vec{k},t)
       \mathbb{S}_{ij}(\vec{k},\vec{x}) \cr&\qquad{}
     + {1 \over k} \partial_t q^{\rm even}_V(\vec{k},t)
       \mathbb{V}^{\rm even}_{ij}(\vec{k},\vec{x})
     + {1 \over k} \partial_t q^{\rm odd}_V(\vec{k},t)
       \mathbb{V}^{\rm odd}_{ij}(\vec{k},\vec{x})  \cr&\qquad{}
     + q_T^{\rm even}(\vec{k},t)
     \mathbb{T}_{ij}^{\rm even}(\vec{k},\vec{x}) +
    q_T^{\rm odd}(\vec{k},t)
     \mathbb{T}_{ij}^{\rm odd}(\vec{k},\vec{x}) \bigg] \,.
 }
The first term on the right hand side of \eno{TaiExpand} must take the form indicated in order to have $\partial^m \langle T_{m0} \rangle_{\rm conserved} = 0$.  The first term on the right hand side of \eno{TijExpand} is dictated by the conformal condition $\langle T^m{}_m \rangle_{\rm conserved} = 0$, and the next three terms are consequences of the conservation equations $\partial^m \langle T_{mi} \rangle_{\rm conserved} = 0$.

The boundary asymptotics of each master field encodes the value of the corresponding $q$ coefficient in \eno{TabExpand}--\eno{TijExpand}, as we explain more precisely in sections~\ref{TENSOR_LIMITS}--\ref{SCALAR_LIMITS}.  We have already remarked that the odd perturbations vanish, and so in the following we omit parity labels.  A subtlety in the expansions \eno{TabExpand}--\eno{TijExpand} is that they actually {\it don't} capture the full stress tensor, which is
 \eqn{TotalTmn}{
  \langle T_{mn} \rangle = \langle T_{mn} \rangle_{\rm bath} + \langle T_{mn} \rangle_{\rm conserved} + \langle T_{mn} \rangle_{\rm drag}\,.
 }
Here $\langle T_{mn} \rangle_{\rm bath} = {\pi^2 \over 8} N^2 T^4 \diag\{3,1,1,1\}$ is the stress-energy dual to the black hole in $AdS_5$-Schwarzschild.  The last term is not conserved and is due to an external force required to balance the drag force of \cite{Herzog:2006gh,Gubser:2006bz}.  The presence of this term was noted in \cite{Friess:2006fk}, and we rederive it in section~\ref{QUARK_FORCE}.

In order to compute $\langle T_{mn} \rangle$ using AdS/CFT, the most straightforward approach is to express the perturbation \eno{GmunuSeries} in axial gauge, $h_{r\mu} = 0$.  As will be explained in detail below, the non-zero components of the metric perturbations have the following large $r$ behavior:\footnote{The quantities $P_{mn}$, $Q^{\rm tot}_{mn}$, and $R_{mn}$ differ from the corresponding quantities in \cite{Friess:2006fk} by different normalization factors.}
 \eqn{MetricBdyAsymp}{
  h_{mn} = R_{mn} r^2 + P_{mn} {1\over r} + Q^{\rm tot}_{mn} {1\over r^2} + {\cal O} (r^{-3})\,,
 }
where $R_{mn}$, $P_{mn}$, and $Q^{\rm tot}_{mn}$ may depend on $x^m = (t,\vec{x})$ but not on $r$.  Up to a subtlety regarding residual gauge freedom,\footnote{The subtlety is that the boundary condition $R_{mn}=0$ is actually a little more restrictive than necessary.  It should be allowed to have $R_{mn} \propto g_{mn}$, corresponding to adding $T^m{}_m=0$ to the lagrangian, or $R_{mn} \propto v_{(m} k_{n)}$, corresponding to adding a multiple of $\partial_m T^{mn}$ to the lagrangian.  Such deformations correspond to metric perturbations in $AdS_5$-Schwarzschild which are pure gauge.} the boundary condition $R_{mn}=0$ corresponds to requiring that the metric on the boundary should remain fixed, or in other words not adding components of $T_{mn}$ to the gauge theory lagrangian.  Another boundary condition comes from requiring purely infalling modes at the black hole horizon.

Because $h_{mn}$ is a perturbation around $AdS_5$-Schwarzschild rather than pure $AdS_5$, it relates only to the second and third terms in \eno{TotalTmn}.  More precisely, as argued in \cite{Friess:2006fk}, one has
 \eqn{GotTmn}{
  \langle T_{mn} \rangle_{\rm tot} \equiv
    \langle T_{mn} \rangle_{\rm conserved} +
     \langle T_{mn} \rangle_{\rm drag} =
     {2 \over \kappa_5^2 L^3} Q_{mn}^{\rm tot} \,.
 }
The notation $\langle T_{mn} \rangle_{\rm tot}$ is meant to indicate the total VEV arising from the trailing string: it excludes $\langle T_{mn} \rangle_{\rm bath}$.  Actually, it excludes one other thing: the $P_{mn}/r$ term in \eno{MetricBdyAsymp} signals the need for a counter-term subtraction to make $\langle T_{mn} \rangle$ well-defined.  As discussed in \cite{Friess:2006fk}, this subtraction is a delta-function supported at the location of the quark.

The main purpose of sections~\ref{TENSOR_LIMITS}--\ref{SCALAR_LIMITS} is to relate the boundary asymptotics of solutions to the radial master equations \eno{psiIEQ} to the quantities $Q_{mn}^{\rm tot}$ entering \eno{MetricBdyAsymp}.  In so doing we will discover a natural split of $Q_{mn}^{\rm tot}$ into a non-conserved piece which can be computed analytically and a conserved piece which must in general be obtained by solving the master equations with appropriate boundary conditions.  We also explain what these boundary conditions are, thereby completing, in the master field formalism, a specification of the boundary value problem that determines the holographic stress tensor.

\subsection{The asymptotics of the tensor master equation}
\label{TENSOR_LIMITS}

Near the boundary of $AdS_5$-Schwarzschild, the solution to the tensor master equation behaves as
 \eqn{TensorAsympBdy}{
  \psi_T(r) = R_T \left( 1 - {\alpha \over 4 r^2} - {\alpha^2 \over 16 r^4}\log r \right) + P_T {1\over r^3} + Q^{\rm tot}_T {1\over r^4} + {\cal O}(r^{-5}) \,,
 }
where $P_T$ and $\alpha$ are given by
 \eqn{PTDef}{
  P_T = {\ell_v\over \kappa_5^2} {v^2 k_\perp^2 \over 3 k^2} \qquad \alpha \equiv  L^4 (k^2 - v^2 k_1^2)
 }
and $Q^{\rm tot}_T$ and $R_T$ are the two integration constants.  Since $H_T^{T, {\rm even}} = {1\over 2} {\kappa_5^2 L^5} \psi_T(r) e^{-i v k_1 t}$, the requirement $R_{mn} = 0$ translates into $R_T = 0$.  Assuming $R_T = 0$, we have
 \eqn{IKVarsTensor}{
  H_T^{T, {\rm even}} =  \ell_v {v^2 k_\perp^2 \over 6 k^2}  e^{-i v k_1 t} {L^5 \over r^3} + {\kappa_5^2\over 2 L} Q^{\rm tot}_T e^{-i v k_1 t} {L^6 \over r^4} + {\cal O}(r^{-5})\,,
 }
which can be used to find the tensor components of the $Q^{\rm tot}_{mn}$ coefficients via \eqref{hbcDecomp}--\eqref{hijDecomp}.  Feeding the result through \eno{GotTmn} and comparing to the expansion \eno{TijExpand}, one finds that
 \eqn{GotqtEven}{
  q_T(\vec{k},t) = Q_T^{\rm tot}(\vec{k}) e^{-ivk_1 t} \,.
 }
(Recall that because odd parity perturbations vanish for the trailing string, we are always referring to even parity; however, for a more complicated gravitational source, \eno{GotqtEven} could be used equally in reference to odd parity modes.)  The normalization factor between $\Phi_T$ and $\psi_T$ in \eno{RedefineMaster} was chosen to make the relation \eno{GotqtEven} simple.

Near the horizon, the leading behavior of $\psi_T$ is
 \eqn{TensorAsympHor}{
  \psi_T(r) = U_T \left(r - r_H \right)^{-i v k_1 L^2/4 r_H} + V_T \left(r - r_H \right)^{i v k_1 L^2/4 r_H} + {\cal O}(r-r_H)\,,
 }
and the infalling boundary condition is $V_T = 0$.

\subsection{The asymptotics of the vector master equation}
\label{VECTOR_LIMITS}

In the vector case, the master field has the following asymptotic behavior at large $r$:
 \eqn{VectorAsympBdy}{
  \psi_V(r) = R_V {1\over r} \left(1 - {\alpha \over 2 r^2}\log r \right)+ P_V {1\over r^2} + Q^{\rm tot}_V {1\over r^3} + {\cal O}(r^{-4}) \,,
 }
where
 \eqn{PVDef}{
  P_V = -{\ell_v\over \kappa_5^2} {v k_\perp \over k} \qquad \alpha \equiv  L^4 (k^2 - v^2 k_1^2)
 }
and $Q^{\rm tot}_V$ together with $R_V$ are integration constants.  Relating this large $r$ series expansion to \eqref{MetricBdyAsymp} is not hard:  one can use the definition \eqref{RedefineMaster} and plug \eqref{VectorAsympBdy} into \eqref{FaVSmaster} to obtain a large $r$ asymptotic expressions for the gauge-invariant quantities $\hat{f}_b$, and then use the expression for $\hat{f}_b$ in \eqref{FaDef} as well as the axial gauge condition $f_r^{V} = 0$ to find large $r$ series expansions for both $f_t^{V}$ and $H_T^{V}$.  When the latter quantities are plugged into \eqref{hbcDecomp}--\eqref{hijDecomp}, one can easily check that all non-zero components of $R_{mn}$ turn out to be proportional to $R_V$, so setting $R_{mn} = 0$ means setting $R_V = 0$.

When $R_V=0$, the large $r$ behavior of the functions that enter in the metric perturbations is given in axial gauge by
 \eqn{IKVarsVector}{
  f_t^{V, {\rm even}} &= -\ell_v {2 v k_\perp \over 3 k} e^{-i v k_1 t} {L^4\over r^2} + {\kappa_5^2\over 2 L} \left( Q^{\rm tot}_V - {\ell_v\over \kappa_5^2} {i r_H^2 k_\perp\over k k_1 L^2} \right) e^{-i v k_1 t} {L^5\over r^3} + {\cal O} (r^{-4})\cr
  H_T^{V, {\rm even}} &= \ell_v {2 i v^2 k_1 k_\perp \over 3 k^2} e^{-i v k_1 t} {L^5\over r^3} - {\kappa_5^2\over 2 L} {i v k_1 \over k} Q^{\rm tot}_V e^{-i v k_1 t} {L^6\over r^4} + {\cal O} (r^{-5})\,.
 }
Feeding \eno{IKVarsVector} through \eno{GotTmn} and comparing to \eno{TaiExpand}, one finds that
 \eqn{GotqvEven}{
  q_V(\vec{k},t) = Q^{\rm tot}_V(\vec{k}) e^{-i v k_1 t} \,,
 }
provided one identifies
 \eqn{TaiDrag}{
  \langle T_{0i} \rangle_{\rm drag} &=
   - {\ell_v\over \kappa_5^2} {i r_H^2 \over L^2} \int {d^3 k \over (2\pi)^3}
     {k_\perp\over k k_1} e^{-i v k_1 t}
      \mathbb{V}_i^{\rm even}(\vec{k},\vec{x}) \,.
 }

At the horizon, the situation is similar to the one encountered in the tensor case.  The leading behavior of $\psi_V$ is
 \eqn{VectorAsympHor}{
  \psi_V(r) = U_V \left(r - r_H \right)^{-i v k_1 L^2/4 r_H} + V_V \left(r - r_H \right)^{i v k_1 L^2/4 r_H} + {\cal O}(r-r_H)\,,
 }
and again the infalling boundary condition is $V_V = 0$.

\subsection{The asymptotics of the scalar master equation}
\label{SCALAR_LIMITS}

Near the boundary, the asymptotic behavior of $\psi_S(r)$ obeying \eqref{psiIEQ} is
 \eqn{ScalarAsympBdy}{
  \psi_S(r) = P_S {1\over r} + R_S {1\over r^2} \log r + Q^{\rm tot}_S {1\over r^2} + {\cal O}(r^{-3}) \,,
 }
with $P_S$ explicitly given by
 \eqn{PSDef}{
  P_S = {\ell_v\over \kappa_5^2} {k^2 (2 + v^2) - 3 v^2 k_1^2  \over 2 k^2}\,,
 }
and $Q^{\rm tot}_S$ and $R_S$ being regarded as the two integration constants of the second order equation \eqref{psiIEQ}.

The asymptotic behavior of $\psi_S(r)$ given above can be related to the asymptotic behavior of the metric perturbations \eqref{MetricBdyAsymp} through a straightforward but tedious calculation:  one just has to use \eqref{RedefineMaster} to go back to the master field $\Phi_S$ and then trace back the steps outlined in section~\ref{SCALARMASTER} in the special case where the background is $AdS_5$-Schwarzschild.  It is important to note that in doing so we only need to consider large $r$ series solutions to the differential equations we encounter---a much easier task than solving these equations for all $r$.  We will now provide an outline of the computation.

First, one can easily obtain expressions for $\hat{f}^S_{bc}$ and $\hat{H}_L^S$ in terms of $\psi_S(r)$ that are valid for all $r$ from combining the definitions \eqref{HLSelim} and \eqref{ExpressFabS} with equations \eqref{FullXYZ} and with the definition \eqref{MasterSDef} of the master field.  Next, one can find large $r$ asymptotic expansions for $\hat{f}^S_{bc}$ and $\hat{H}_L^S$ by replacing $\psi_S(r)$ with a large $r$ series solution to the master equation.  While the first few terms of this solution are given in \eqref{ScalarAsympBdy}, an improved series expansion (up to order ${\cal O}\left(r^{-8}\right)$) is needed to arrive at the results listed below.  The last step of the computation is to assume axial gauge, namely $f_r^S = f_{tr}^S = f_{rr}^S = 0$, and to use the asymptotic expansions for $\hat{f}^S_{bc}$ and $\hat{H}_L^S$ to find series solutions for the quantities $f_{tt}^S$, $f_t^S$, $H_T^S$, and $H_L^S$ that obey the system of differential equations \eqref{ScalarInvariants}.  The asymptotics for the non-zero metric perturbations $h_{mn}$ can then be found from \eqref{hbcDecomp}--\eqref{hijDecomp}, and one can check they have the form \eqref{MetricBdyAsymp}.  Moreover, all scalar components of $R_{mn}$ end up being proportional to $R_S$, so again $R_{mn} = 0$ means $R_S = 0$.

Assuming $R_S = 0$, the expressions one ends up with for the axial gauge metric perturbations are
\begin{align}
  f_{tt}^S &= \ell_v {2 (2 + v^2)\over 9} e^{-i v k_1 t} {L^3\over r} + {\kappa_5^2\over 2 L}\left(Q^{\rm tot}_S +  {\ell_v\over \kappa_5^2} {i v r_H^2 \over k_1 L^2 } \right) e^{-i v k_1 t} {L^4\over r^2} + {\cal O} (r^{-3}) \notag\\
  f_t^S &= -\ell_v {2 i v k_1\over 3 k} e^{-i v k_1 t} {L^4\over r^2} - {\kappa_5^2\over 2 L} {i v k_1 \over k} Q^{\rm tot}_S e^{-i v k_1 t} {L^5\over r^3} + {\cal O} (r^{-4}) \notag \\
  H_L^S &= \ell_v {1\over 9} e^{-i v k_1 t} {L^5\over r^3} + {\kappa_5^2\over 2 L} \left( {1\over 6} Q^{\rm tot}_S + {\ell_v\over \kappa_5^2} {i v r_H^2\over 6 k_1 L^2} \right)e^{-i v k_1 t} {L^6\over r^4} + {\cal O} (r^{-5}) \label{IKVarsScalar}\\
  H_T^S &= \ell_v {v^2 (k^2 - 3 k_1^2)\over 6 k^2} e^{-i v k_1 t} {L^5\over r^3} + {\kappa_5^2\over 2 L} \biggl( {k^2 - 3 v^2 k_1^2\over 4 k^2} Q^{\rm tot}_S \notag\\
  &\hspace{2.1in}+ {\ell_v\over \kappa_5^2} {i r_H^2 v (k^2 + 3 k_1^2) \over 4 k^2 k_1 L^2} \biggr)e^{-i v k_1 t} {L^6\over r^4} + {\cal O} (r^{-5})\,. \notag
\end{align}
As in the tensor and vector cases, feeding \eno{IKVarsScalar} through \eno{GotTmn} and comparing to \eno{TabExpand}, one finds that
 \eqn{GotqsEven}{
  q_S(\vec{k},t) = Q^{\rm tot}_S(\vec{k}) e^{-i v k_1 t} \,,
 }
provided one identifies
 \eqn{TabDrag}{
  \langle T_{00} \rangle_{\rm drag} = {\ell_v\over \kappa_5^2} {i v r_H^2 \over L^2 } \int {d^3 k \over (2\pi)^3}
     {1\over k_1} e^{-i v k_1 t} \mathbb{S}(\vec{k},\vec{x})
 }
 \eqn{TijDrag}{
  \langle T_{ij} \rangle_{\rm drag} &=
   2 {\ell_v\over \kappa_5^2} {i r_H^2 \over L^2} \int {d^3 k \over (2\pi)^3}  e^{-i v k_1 t}
    \left(  {1 \over 6 k_1} \delta_{ij}
       \mathbb{S}(\vec{k},\vec{x}) + {k^2 + 3 k_1^2 \over 4 k^2 k_1}
       \mathbb{S}_{ij}(\vec{k},\vec{x}) \right) \,.
 }
From \eno{IKVarsScalar} one may also find explicit expressions for the quantities $P_{mn}$ appearing in \eno{MetricBdyAsymp}.  Using \eqref{TVSToJuly} in appendix~\ref{JULYVARIABLES} one can check that the expressions so derived match with the corresponding $P_{mn}$ coefficients calculated in \cite{Friess:2006fk}.

At the horizon, the series expansion for $\psi_S$ is
 \eqn{ScalarAsympHor}{
  \psi_S(r) = U_S \left(r - r_H \right)^{-i v k_1 L^2/4 r_H} + V_S \left(r - r_H \right)^{i v k_1 L^2/4 r_H} + {\cal O}(r-r_H)\,.
 }
The infalling boundary condition is $V_S = 0$.

\subsection{The force on the moving quark}
\label{QUARK_FORCE}

In sections~\ref{VECTOR_LIMITS} and~\ref{SCALAR_LIMITS}, we saw the non-conserved part of the stress tensor, $\langle T_{mn} \rangle_{\rm drag}$, emerging from mismatches between the Fourier expansion \eno{TabExpand}--\eno{TijExpand} of a general conserved, traceless stress tensor and the boundary asymptotics of the metric perturbation $h_{mn}$, as computed by tracing backward through the master field definitions in sections~\ref{VECTORMASTER} and~\ref{SCALARMASTER}.  In this section we recheck an observation of \cite{Friess:2006fk}, namely that the non-conservation of $\langle T_{mn} \rangle_{\rm drag}$ is the result of an external force on the quark that precisely cancels the drag force so as to keep the quark moving at constant velocity.  Equivalently, one can regard this non-conservation of $\langle T_{mn} \rangle_{\rm drag}$ as describing the rate at which energy and momentum are supplied to the bath by the external quark.

The external force on the quark is
 \eqn{ForceDef}{
  F^i = \int_V d^3 x \, \partial_m \langle T^{mi} \rangle_{\rm drag}\,,
 }
where the integration volume $V$ is any three-dimensional volume enclosing the quark.  The quantity $\partial_m \langle T^{mn} \rangle$ can be easily calculated from \eno{TaiDrag}, \eno{TabDrag}, and~\eno{TijDrag}:
 \eqn{ConservQmn}{
  \partial_m \langle T^{mn} \rangle = {\ell_v\over \kappa_5^2} {v r_H^2 \over L^2} \delta(x^1-vt) \delta(x^2) \delta(x^3) \begin{pmatrix} v & 1 & 0 & 0 \end{pmatrix}\,.
 }
Evaluating the integral and using $\ell_v = \kappa_5^2 / 2 \pi \alpha' \sqrt{1-v^2}$, we obtain
 \eqn{MinusDrag}{
  F^1 = {v r_H^2\over 2 \pi \alpha' L^2 \sqrt{1-v^2}} = {\pi \sqrt{g_{YM}^2 N}\over 2} T^2 {v\over \sqrt{1-v^2}}\,,
 }
which is indeed minus the drag force \cite{Herzog:2006gh, Gubser:2006bz}.  In obtaining the last equality we have used $T = r_H/\pi L^2$ as well as $L^2/\alpha' = \sqrt{g_{YM}^2 N}$.

\section{Small $k$ approximation}

If $k \ll T$, the equations for the $\psi_I$ simplify and can be solved analytically.  In fact, we can use the regular perturbation theory technique to find small $k$ series solutions for each of these equations.  The series solutions will in turn enable us to extract small $k$ series expansions for $Q^{\rm tot}_T$, $Q^{\rm tot}_V$, and $Q^{\rm tot}_S$, which are needed to find the small $k$ behavior of $\langle T_{mn} \rangle$.

\subsection{Tensor set}
\label{SMALLKTENSOR}

As remarked before, the tensor equation is the same as the dilaton equation in \cite{Friess:2006aw}, or the $A$-equation in \cite{Friess:2006fk}.  While the derivation of the small $k$ solution of this equation can be also found in section~3.5 of \cite{Friess:2006fk}, we will now outline the main steps of the calculation.

We start with the equation for $\psi_T$ in \eqref{psiIEQ} and develop a series expansion in $k$:
 \eqn{PsiSmallKTensor}{
  \psi_T (r) = \psi_0(r) + k \psi_1(r) + k^2 \psi_2(r) + \ldots \,.
 }
The equations satisfied by each $\psi_i$ can be found from equating like powers of $k$ in \eqref{psiIEQ}.  In doing so, we consider $k_1$ and $k_\perp$ to be of the same order of smallness as $k$, so for example a term containing $k_1 k_\perp k^2$ will be considered to be ${\cal O} (k^4)$.  These equations can then be solved by requiring that the boundary and horizon asymptotics of the solutions be nothing but small $k$ expansions of \eqref{TensorAsympBdy} and \eqref{TensorAsympHor} with $R_T = 0$ and $V_T = 0$.  For example, the equation for $\psi_0$ is
 \eqn{psi0TEQ}{
  \left[\partial_r^2 + \left({5\over r} + {h'\over h}\right)\partial_r \right]\psi_0 = - {\ell_v\over \kappa_5^2} {v^2 k_\perp^2 \over k^2 r h} \,,
 }
and its solution is
 \eqn{psi0Soln}{
  \psi_0 (r)= {\ell_v\over \kappa_5^2} {v^2 k_\perp^2\over 4 k^2 r_H^3} \left(2 \tan^{-1} {r\over r_H} - \pi + \log {r+r_H\over r-r_H} + \log{r^4 - r_H^4\over r^4} \right)\,.
 }
Expanding this solution in series at large $r$, we obtain
 \eqn{psi0Far}{
  \psi_0(r) = {\ell_v\over \kappa_5^2} {v^2 k_\perp^2\over 3 k^2} {1\over r^3} - {\ell_v\over \kappa_5^2} {v^2 k_\perp^2 r_H\over 4 k^2} {1\over r^4} + {\cal O} (r^{-5})\,,
 }
from which we can identify the coefficient of $1/r^4$ with the leading contribution to $Q_T^{\rm tot}$.

Performing a similar analysis for $\psi_1$ and $\psi_2$, we find
 \eqn{QTSmallK}{
  Q^{\rm tot}_T &= -{\ell_v\over \kappa_5^2} {r_H v^2 k_\perp^2\over 4 k^2}  + {\ell_v\over \kappa_5^2} {i v^3 k_1 k_\perp^2 L^2 \log 2 \over 8 k^2 } \cr&\qquad\qquad{} + {\ell_v\over \kappa_5^2} {v^2 k_\perp^2 L^4 (6 \pi - 12 \log 2)\over 192 k^2 r_H} \left[k^2 - v^2 k_1^2  + v^2 k_1^2 {\pi^2 - 12 (\log 2)^2\over 6 \pi - 12 \log 2} \right] + {\cal O}(k^3)\,.
 }

\subsection{Vector set}
\label{SMALLKVECTOR}

The small $k$ analysis in the vector goes mostly the same way as in the tensor case:  one starts by expanding $\psi_V$ in a low-$k$ series
 \eqn{PsiSmallKVector}{
  \psi_V (r) = \psi_0(r) + k \psi_1(r) + k^2 \psi_2(r) + \ldots \,,
 }
and then the equations for $\psi_i$ follow from equating like powers of $k$ in the expansion of the $\psi_V$ equation in \eqref{psiIEQ}.  The difference from the tensor case is that the matching of the $\psi_i$ to the asymptotics \eqref{VectorAsympBdy} and \eqref{VectorAsympHor} with $R_V = V_V = 0$ and appropriately expanded in $k$ results in one undetermined integration constant at each order $i$.  Fortunately, this integration constant can be determined simply by looking at the horizon asymptotics of $\psi_{i+1}$.  For example, $\psi_0$ is given exactly by
 \eqn{psi0Vector}{
  \psi_0(r) = -{\ell_v \over \kappa_5^2}  {v k_\perp (r-r_H)\over k r^3} + U_0 {r_H^3\over r^3}\,,
 }
where $U_0$ is an integration constant of order ${\cal O}(k^0)$.  It is not hard to see that, to order ${\cal O}(k^0)$, this expression satisfies the required boundary conditions:  $\psi_0$ looks like an infalling wave at $r=r_H$ and there's no $1/r$ piece at large $r$ regardless of the value of $U_0$.  The only way to determine $U_0$ is to go to order ${\cal O}(k)$.

We obtain the following small $k$ expansion for $Q^{\rm tot}_V$:
 \eqn{QVSmallK}{
  Q^{\rm tot}_V = {\ell_v\over \kappa_5^2} {r_H k_\perp (4 v^2 k_1^2-k^2)\over 4 v k k_1^2} + {\ell_v\over \kappa_5^2} {i k_\perp  L^2 \left[k^4 - (8\log 2) v^4 k_1^4  + (2 \log 2)  v^2 k^2 k_1^2\right] \over 16 v^2 k k_1^3 } +{\cal O} (k^2)\,.
 }

\subsection{Scalar set}
\label{SMALLKSCALAR}

Obtaining a small $k$ solution for the scalar equation in \eqref{psiIEQ} is a bit tricky, because the quantity $\psi_S$ doesn't have a small $k$ expansion itself.  Instead, we can find a small $k$ approximation for the rescaled field
 \eqn{psitildeDef}{
  \tilde\psi_S(r) = (k^2 + {6 r_H^4\over L^4 r^2}) \psi_S(r)\,.
 }
To do so, we can follow the same steps as the ones outlined in the tensor case in section~\ref{SMALLKTENSOR} and replace $\psi_T$ by $\tilde\psi_S$.  It is not hard to check that the zeroth order solution $\psi_0$ is zero in this case, so the first non-trivial term should be first order in $k$:
 \eqn{PsiSmallKScalar}{
  \tilde\psi_S (r) = k \psi_1(r) + k^2 \psi_2(r) + k^3 \psi_3(r) + \ldots \,.
 }
One has to be careful though when matching the solutions for $\psi_i$ to the asymptotics \eqref{ScalarAsympBdy} and \eqref{ScalarAsympHor}:  the rescaling factor in \eqref{psitildeDef} reduces to $k^2$ in the large $r$ limit, so the order of the $\tilde\psi_S$ expansion changes by two when we look at the boundary asymptotics.  With this minor technical difference, the matching proceeds in the same way as in the tensor case.

By reading off the coefficient of the $1/r^2$ term in the large $r$ expansion of $\tilde\psi_S$, we find $Q^{\rm tot}_S$ to be
 \eqn{QSSmallK}{
  Q^{\rm tot}_S = -{\ell_v\over \kappa_5^2} {i v r_H^2 (k^2 + 3 k_1^2)\over L^2 k_1 (k^2 - 3 v^2 k_1^2)} + {\ell_v\over \kappa_5^2} {3 v^2 r_H k_1^2 \left[-3 v^2 k_1^2 + k^2 (2 + v^2) \right]\over (k^2 - 3 v^2 k_1^2)^2} + {\cal O}(k)\,.
 }
This formula diverges at the Mach angle, namely where the ratio of the momenta is $k_1/k = 1/v\sqrt{3}$, signaling a sonic boom in the thermal medium.  While the divergence appears to get worse as one goes to higher orders in the expansion, the singular behavior disappears if the series expansion \eqref{QSSmallK} is resummed into a form that resembles a Lorentzian lineshape.  For a more detailed discussion, see section~3.5 of \cite{Friess:2006fk}.

\section{Large $k$ approximation}
\label{LARGEK}

In \cite{Yarom:2007ap}, a method of Green's functions was employed to determine the large $k$ asymptotic form of the VEV of the operator ${\cal O}_\phi$ dual to the dilaton, as explored numerically in \cite{Friess:2006aw}.  This method can be understood as an expansion in powers of $r_H$, which makes sense because $r_H \propto T$ and all $k$-dependence of dimensionless quantities arises in series expansions in $k/T$.  Suppose one starts with an ordinary differential equation of the form
 \eqn{LpsiJ}{
  {\cal L} \psi(r) = \left[ \partial_r^2 + t(r) \partial_r + u(r)
    \right] \psi(r) = -j(r)
 }
where there is an expansion
 \eqn{rHexpand}{
  {\cal L} &= {\cal L}_0 + r_H {\cal L}_1 + r_H^2 {\cal L}_2 +
    \ldots  \cr
  \psi &= \psi_0 + r_H \psi_1 + r_H^2 \psi_2 + \ldots  \cr
  j &= j_0 + r_H j_1 + r_H j_2 + \ldots \,.
 }
The radial master equations \eno{psiIEQ} fit this form after both sides are divided by $f = (1-r_H^4/r^4) r^2/L^2$.  At least formally, one may solve \eno{LpsiJ} order-by-order in $r_H$.  At the lowest order, ${\cal L}_0$ is a differential operator in pure $AdS_5$, which is the $r_H \to 0$ limit of $AdS_5$-Schwarzschild.  Construct two solutions of the homogeneous equation,
 \eqn{psiHpsiVEV}{
  {\cal L}_0 \psi_H = 0 = {\cal L}_0 \psi_{\rm VEV} \,,
 }
where $\psi_H$ satisfies appropriate boundary conditions at the degenerate horizon $r=0$, while $\psi_{\rm VEV}$ has appropriate falloff at the boundary of $AdS_5$ to describe a VEV rather than a deformation.  Then a Green's function inverse to ${\cal L}_0$, satisfying
 \eqn{GreenDef}{
  {\cal L}_0 G_0(r;r_0) = \delta(r-r_0)
 }
as well as the boundary conditions at the horizon and infinity satisfied by $\psi_H$ and $\psi_{\rm VEV}$, respectively, may be constructed as
 \eqn{GreenEq}{
  G_0(r;r_0) = {1 \over W[\psi_H,\psi_{\rm VEV}](r_0)}
    \left\{ \seqalign{\span\TL & \qquad\span\TT}{
     \psi_H(r_0) \psi_{\rm VEV}(r) & for $r>r_0$  \cr
     \psi_{\rm VEV}(r_0) \psi_H(r) & for $r<r_0$
    } \right.
 }
where the Wronskian is
 \eqn{WronDef}{
  W[\psi_H,\psi_{\rm VEV}](r_0) = \psi_H(r_0) \psi_{\rm VEV}'(r_0) -
    \psi_{\rm VEV}(r_0) \psi_H'(r_0) \,.
 }
The leading order solution $\psi_0$ may be determined as follows:
 \eqn{DeterminePsi}{
  \psi_0(r) &= -(G_0 * j_0)(r) \equiv
    -\int_0^\infty dr_0 \, G_0(r;r_0) j_0(r_0)  \cr
    &= A_0(r) \psi_H(r) + B_0(r) \psi_{\rm VEV}(r)
 }
where
 \eqn{AzeroBzero}{
  A_0(r) = -\int_r^\infty dr_0
        {\psi_{\rm VEV}(r_0) \over W(r_0)} j_0(r_0) \qquad
  B_0(r) = -\int_0^r dr_0
        {\psi_H(r_0) \over W(r_0)} j_0(r_0) \,.
 }
In optimal circumstances, $B_0(\infty) = \lim_{r\to\infty} B_0(r)$ is finite and $A_0(r) \to 0$ fast enough as $r \to 0$ so that the boundary asymptotics of $\psi_0(r)$ may be expressed as
 \eqn{OptimalBC}{
  \psi_0(r) = B_0(\infty) \psi_{\rm VEV}(r) + \hbox{(sub-leading)} \,.
 }
The form \eno{OptimalBC} is convenient because, up to a normalizing factor, $B_0(\infty)$ is just the ${\cal O}(r_H^0)$ contribution to $Q^{\rm tot}_T$, $Q^{\rm tot}_V$, or $Q^{\rm tot}_S$.  Unfortunately, there is a complication in the cases we will study: $B_0(r)$ does not have a finite limit.  The solution is to isolate a simple analytic form $\psi_0^{\rm CT}(r)$, capturing that part of $\psi_0(r)$ whose boundary asymptotics are more singular than indicated in \eno{OptimalBC}, and split both $\psi_0$ and $j_0$ so that
 \eqn[c]{psiSplit}{
  \psi_0 = \psi_0^{\rm CT} + \psi_0^{\rm R} \qquad
   j_0 = j_0^{\rm CT} + j_0^{\rm R}  \cr
  {\cal L}_0 \psi_0^{\rm CT} = -j_0^{\rm CT} \qquad
   {\cal L}_0 \psi_0^{\rm R} = -j_0^{\rm R} \,.
 }
One may then follow through the Green's function analysis of the ``renormalized'' equation ${\cal L}_0 \psi_0^{\rm R} = -j_0^{\rm R}$ as indicated in \eno{DeterminePsi}--\eno{AzeroBzero} and find that the ``renormalized'' analog of \eno{OptimalBC} holds with finite $B_0^{\rm R}(\infty)$.  The splitting \eno{psiSplit} is in the spirit of holographic renormalization as reviewed for example in \cite{Skenderis:2002wp}, although it also follows an older and more elementary line of thinking where one looks for leading non-analytic contributions to Green's functions in momentum space: see for example \cite{Gubser:1998bc}.

Higher-order $\psi_i$ may be determined in a similar fashion:
 \eqn{HigherOrders}{
  \psi_1(r) &= -(G_0 * \tilde{j}_1)(r) \qquad
    \tilde{j}_1 = j_1 + {\cal L}_1 \psi_0  \cr
  \psi_2(r) &= -(G_0 * \tilde{j}_2)(r) \qquad
    \tilde{j}_2 = j_2 + {\cal L}_2 \psi_0 + {\cal L}_1 \psi_1  \cr
  \ldots \,.
 }
A ``renormalization'' process similar to \eno{psiSplit} could be employed, if needed, at higher orders; but need for it would be unexpected given that the ``counter-term'' contributions usually correspond to contact terms in the gauge theory which are temperature-independent.

A significant obstacle to the treatment just described is that the expansion \eno{rHexpand} is not uniform in $r$: near the horizon it breaks down.  It is not clear whether the boundary conditions appropriate to the zero-temperature Green's function $G_0$ lead, through the expansion \eno{rHexpand}, to a solution $\psi$ satisfying standard boundary conditions at the horizon.  This issue was confronted in \cite{Yarom:2007ap} with a WKB treatment.  We will take the less rigorous approach of computing the leading non-trivial $k$-dependence of $Q^{\rm tot}_T$, $Q^{\rm tot}_V$, and $Q^{\rm tot}_S$ and comparing to numerical evaluations from \cite{Friess:2006fk}.  The eventual aim is to give an all-scales description of $\langle T_{mn} \rangle$, both in $k$ space and position space, using a combination of numerical methods for intermediate scales and analytical approximations for the IR and UV limits.

\subsection{Tensor set}
\label{TENSOR_LARGEK}

This is essentially the case considered in \cite{Yarom:2007ap}, because the tensor master equation is the same as the dilaton equation \cite{Friess:2006aw}.  The leading non-zero terms in the expansions\eno{rHexpand} are
 \eqn[c]{LeadingTensor}{
  {\cal L}_0 = \partial_r^2 + {5 \over r} \partial_r -
    {L^4 \over r^4} \tilde{k}^2  \cr
  j_0 = {3 \over r^5} P_T \qquad
  j_2 = {ivk_1 L^2 \over r^8} P_T
 }
where $P_T$ is as defined in \eno{PTDef} and
 \eqn{kTildeDef}{
  \tilde{k}^2 \equiv k^2 - v^2 k_1^2 =
    (1-v^2) k_1^2 + k_\perp^2 \,.
 }
The zero temperature solutions and Wronskian are
 \eqn[c]{ZeroTfcts}{
  \psi_{\rm VEV}(r) = {1 \over r^2} I_2(\tilde{k} L^2/r) \qquad
   \psi_H(r) = {1 \over r^2} K_2(\tilde{k} L^2/r)
     \cr\noalign{\vskip2\jot}
   W(r) = W[\psi_H,\psi_{\rm VEV}](r) = -{1 \over r^5} \,,
 }
where $I_\nu$ and $K_\nu$ denote Bessel functions.  It's significant that $\tilde{k}^2>0$ always because it results in a Green's function that vanishes faster than any power of $r$ as $r \to 0$, i.e.~in the depths of $AdS_5$.\footnote{Probably this means that putting the horizon back at finite $r_H$ corresponds to ${\cal O}(e^{-\tilde{k} L^2/r_H})$ effects for some $c$-number $b_0$.  If so, it means that the series expansion \eno{rHexpand} for $\psi$ in powers of $r_H$ is only an asymptotic series.  Because $r_H/L^2 = \pi T$, one sees that $e^{-\tilde{k} L^2/r_H}$ has essentially the form of a Maxwell-Boltzman factor.}  Using $\psi_0^{\rm CT}(r) = P_T/r^3$, one may straightforwardly show that
 \eqn{FoundPsiZeroTwoT}{
  \psi_0(r) = P_T \left[
   {1 \over r^3} - {3\pi \tilde{k} L^2 \over 16r^4} +
    {\cal O}(r^{-5}) \right] \qquad
  \psi_2(r) = P_T \left[ {ivk_1 \over \tilde{k}^2 L^2 r^4} +
    {\cal O}(r^{-5}) \right] \,.
 }
A ``counter-term'' is not needed for the analysis of $\psi_2(r)$, as expected on general grounds.  Comparing \eno{FoundPsiZeroTwoT} with \eno{TensorAsympBdy} leads immediately to
 \eqn{QTHighK}{
  Q^{\rm tot}_T = P_T \left[ -{3\pi \tilde{k} L^2 \over 16} +
  {ivk_1 \over \tilde{k}^2 L^2} r_H^2 + {\cal O}(r_H^4) \right]
 }

\subsection{Vector set}
\label{VECTOR_LARGEK}

The leading non-zero terms in the expansion \eno{rHexpand} are
 \eqn[c]{LeadingVector}{
  {\cal L}_0 = \partial_r^2 + {5 \over r} \partial_r -
    {L^4 \over r^4} \tilde{k}^2 + {3 \over r^2} \cr
  j_0 = {1 \over r^4} P_V \qquad
  j_2 = {L^2 \over 3 i v k_1 r^7} (3\tilde{k}^2 -
    v^2 k_1^2) P_V
 }
where $P_V$ is as defined in \eno{PVDef}.  The zero temperature solutions and Wronskian are
 \eqn[c]{ZeroVfcts}{
  \psi_{\rm VEV}(r) = {1 \over r^2} I_1(\tilde{k} L^2/r) \qquad
   \psi_H(r) = {1 \over r^2} K_1(\tilde{k} L^2/r)
     \cr\noalign{\vskip2\jot}
   W(r) = W[\psi_H,\psi_{\rm VEV}](r) = -{1 \over r^5} \,.
 }
Using $\psi_0^{\rm CT}(r) = P_V/r^2$ one obtains
 \eqn{FoundPsiZeroTwoV}{
  \psi_0(r) = P_V \left[
   {1 \over r^2} - {\pi \tilde{k} L^2 \over 4r^3} +
    {\cal O}(r^{-4}) \right] \qquad
  \psi_2(r) = P_V \left[ {3\tilde{k}^2-v^2 k_1^2 \over
    3ivk_1 \tilde{k}^2 L^2 r^3} + {\cal O}(r^{-4}) \right]
 }
Comparing to \eno{VectorAsympBdy}, one sees that
 \eqn{QVHighK}{
  Q^{\rm tot}_V = P_V \left[ -{\pi \tilde{k} L^2 \over 4} +
   {3\tilde{k}^2-v^2 k_1^2 \over 3ivk_1 \tilde{k}^2 L^2} r_H^2 +
    {\cal O}(r_H^4) \right] \,.
 }

\subsection{Scalar set}
\label{SCALAR_LARGEK}

The leading non-zero terms in the expansion \eno{rHexpand} are
 \eqn{LeadingScalar}{
  {\cal L}_0 &= \partial_r^2 + {5 \over r} \partial_r -
    {L^4 \over r^4} \tilde{k}^2 + {4 \over r^2}  \cr
  j_0 &= \left[ -{1 \over r^3} + {(2\tilde{k}^2-v^2 k_1^2) L^4
    \over 3r^5} \right] P_S  \cr
  j_2 &= \bigg[ -{2iL^2 v \over 3k_1}
   {3\tilde{k}^4 - 5 v^2 k_1^4 (1-v^2) + k_1^2 \tilde{k}^2
     (1+8v^2) \over
    \tilde{k}^2 (2+v^2) - v^2 k_1^2 (1-v^2)} {1 \over r^6} +
     {ivk_1 \over 9} L^6 r_H^2 {2\tilde{k}^2 - v^2 k_1^2 \over
      r^8} \bigg] P_S
 }
where $P_S$ is as defined in \eno{PSDef}.  The zero temperature solutions and Wronskian are
 \eqn[c]{ZeroSfcts}{
  \psi_{\rm VEV}(r) = {1 \over r^2} I_0(\tilde{k} L^2/r) \qquad
   \psi_H(r) = {1 \over r^2} K_0(\tilde{k} L^2/r)
     \cr\noalign{\vskip2\jot}
   W(r) = W[\psi_H,\psi_{\rm VEV}](r) = -{1 \over r^5} \,.
 }
Using $\psi_0^{\rm CT}(r) = P_S/r$ one obtains
 \eqn{FoundPsiZeroTwoS}{
  \psi_0(r) &= P_S \left[
   {1 \over r} - {\pi L^2 (\tilde{k}^2+v^2 k_1^2) \over
      6\tilde{k} r^2} +
    {\cal O}(r^{-3}) \right]  \cr
  \psi_2(r) &= P_S \left[ -{2iv (\tilde{k}^2 + v^2 k_1^2)
    \over 9k_1 \tilde{k}^4 L^2}
   {9\tilde{k}^4 - 2v^2 (1-v^2) k_1^4 +
       k_1^2 \tilde{k}^2 (-5+11v^2) \over
     (2+v^2) \tilde{k}^2 - v^2 k_1^2 (1-v^2)} +
      {\cal O}(r^{-3}) \right]
 }
Comparing to \eno{ScalarAsympBdy}, one sees that
 \eqn{QSHighK}{
  Q^{\rm tot}_S &= {\ell_v \over \kappa_5^2} \bigg[
    -{\pi L^2 [ (2+v^2) \tilde{k}^2 - v^2 (1-v^2) k_1^2 ] \over
     12 \tilde{k}}  \cr&\qquad\qquad{} +
     {r_H^2 v \over 9L^2} {9\tilde{k}^4 - 2v^2 (1-v^2) k_1^4 +
       k_1^2 \tilde{k}^2 (-5+11v^2) \over
      k_1 \tilde{k}^4} + {\cal O}(r_H^4) \bigg] \,.
 }

The result \eno{QSHighK} is more properly an expansion in $r_H/(L^2 \tilde{k})$, or, for $v$ not too close to $1$, in inverse powers of $K \equiv k L^2/r_H$, which is the dimensionless wave-number used in \cite{Friess:2006fk}.

\begin{figure}
  \includegraphics[width=6.5in]{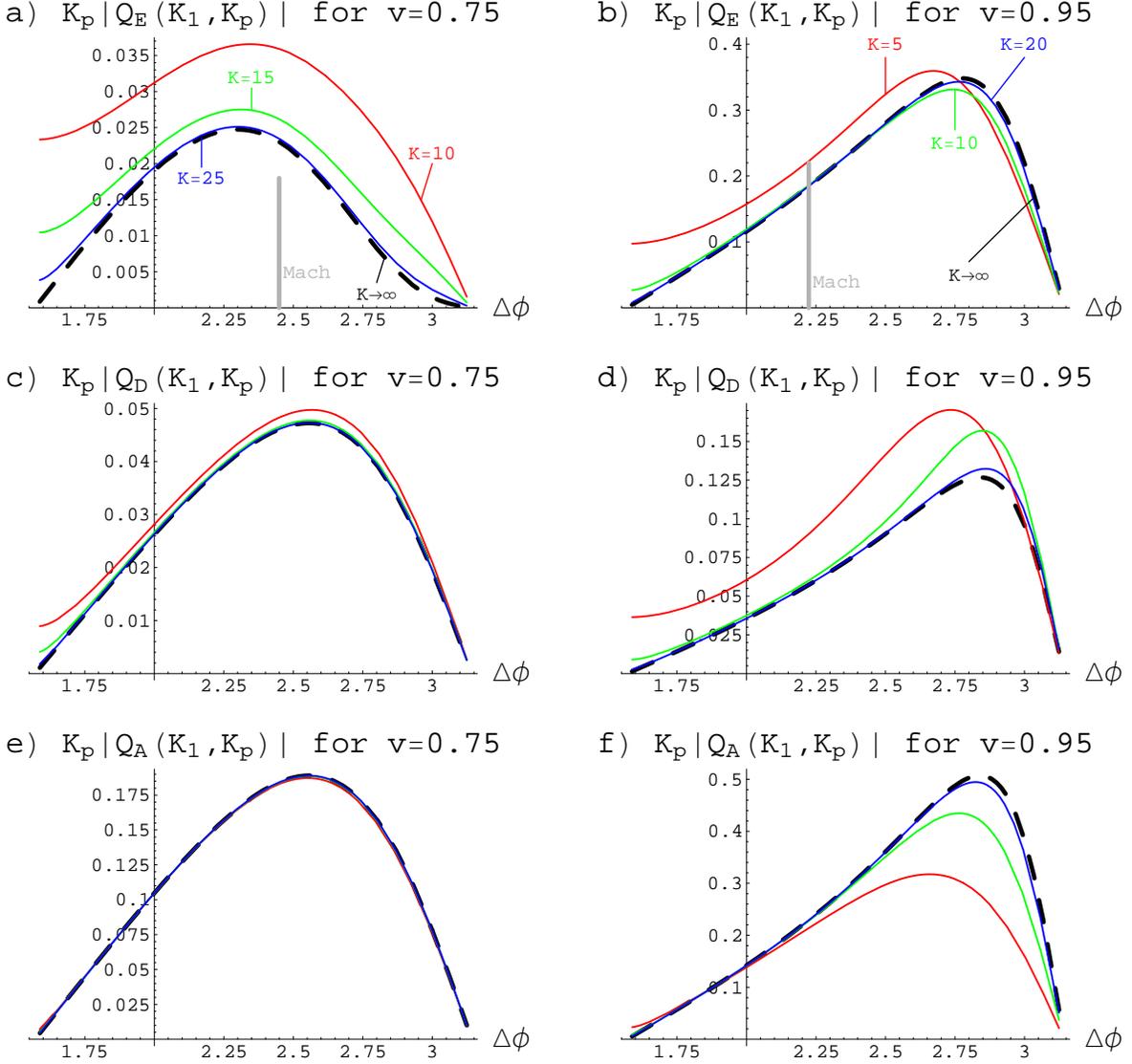}
   \caption{Comparisons between the large $K$ analytic approximations found in \eqref{QSHighK} (plots a,b), \eno{QVHighK} (plots c,d), and~\eno{QTHighK} (plots e,f) and corresponding numerical solutions of the linearized Einstein equations in the form studied in \cite{Friess:2006fk}.  The unbroken colored lines come from numerical solutions at the values of $K=k/\pi T$ indicated, while the dashed lines come from evaluations of \eno{QSHighK}, suitably transformed into evaluations of $Q_E$, $Q_D$, and $Q_A$ as described in the text.  The values of $K$ used in plots c,e are the same as in a, and those in d,f are the same as in b.  For values of $K$ larger than the largest one shown in each plot, the numerical evaluations are almost indistinguishable from the $K\to\infty$ approximation.  The Mach angle is indicated, but there is no reason to expect structure near this angle because large $K$ is the opposite limit of where hydrodynamics is expected to apply.}\label{LargeKmatch}
 \end{figure}

To get an idea of when the next term in the expansion \eno{QSHighK} is significant, we have plotted in figure~\ref{LargeKmatch}a,b a comparison of \eno{QSHighK} with numerical evaluations using the approach of \cite{Friess:2006fk}.  The most immediate conclusion is that the value of $K$ where \eno{QSHighK} becomes reasonably accurate depends significantly on $v$.  To understand the comparison with numerics, first note that we are plotting not $Q_S^{\rm tot}$ but $Q_E$ as defined in \cite{Friess:2006fk}: this quantity is proportional to $Q_S^{\rm tot}$ minus the leading large $k$ behavior in \eno{QSHighK} (corresponding to the pure Coulombic near-field of the quark) plus the non-conserved contribution from \eno{TabDrag}.  The constant of proportionality is the same as the one in \eno{QRelations}.  Also, because of the comparisons with away-side jet-splitting that motivated the study \cite{Friess:2006fk}, we choose for an angular variable not the angle $\theta = \arctan k_\perp/k_1$ between the wave-vector and the direction of the quark's motion, but instead $\Delta\phi = \pi-\theta$.  Thus $\Delta\phi = \pi$ corresponds to wave-vectors exactly along the direction of the quark's motion.  Finally, all dimensionful quantities are rendered dimensionless using factors of $\pi T = r_H/L^2$: for example, $\vec{K} = \vec{k}/\pi T$.

In figure~\ref{LargeKmatch}c,d,e,f we have made corresponding comparisons of the analytic forms \eno{QTHighK} and~\eno{QVHighK} to numerical evaluations following \cite{Friess:2006fk}.  As in the comparison of $Q_S^{\rm tot}$ to $Q_E$, there is a non-trivial affine relation between $Q_V^{\rm tot}$ and the quantity $Q_D$ appearing in the plots, and between $Q_T^{\rm tot}$ and $Q_A$.  These relations can be worked out using \eno{QRelations}, keeping in mind that the quantities plotted exclude the purely Coulombic near-field.

\subsection{The position-space near field}

The large $k$ results from the previous sections can be Fourier transformed to position space without much difficulty.  It is important to remember that $Q^{\rm tot}_T$, $Q^{\rm tot}_V$, and $Q^{\rm tot}_S$ give only the contribution $\langle T_{mn}\rangle_{\rm conserved}$ to the total stress tensor in the plasma, and that the stress tensor of the thermal bath $\langle T_{mn}\rangle_{\rm bath}$ together with a non-conserved piece $\langle T_{mn}\rangle_{\rm drag}$ that is responsible for the drag force need to be added in order to recover the full stress tensor (see equation \eqref{TotalTmn}).  From now on we will focus on the quantity $\langle T_{mn} \rangle_{\rm tot} = \langle T_{mn} \rangle_{\rm conserved} + \langle T_{mn} \rangle_{\rm drag}$, as defined in \eno{GotTmn}.

\begin{figure}
  \includegraphics[width=6.5in]{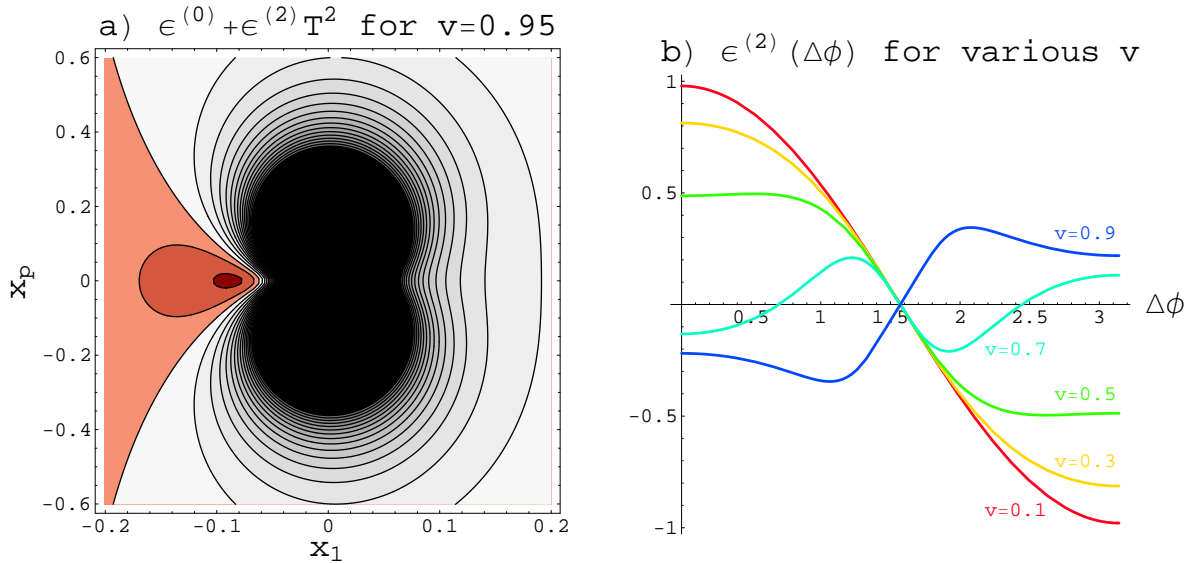}
  \caption{(a) A contour plot of the near-field energy density as computed from \eno{T00Position} at time $t=0$.  The red regions in the left side of the plot represent negative $\langle T_{00} \rangle$---more precisely, an energy deficit compared to the average energy density of the thermal bath.  The deficit arises because the ${\cal O}(T^2)$ dominates over the ${\cal O}(1)$ term, which leads us to question the accuracy of the asymptotics in these regions.  (b) The ${\cal O}(T^2)$ correction $\epsilon^{(2)}$ as a function of $\Delta\phi = \pi - \arctan x_\perp/x_1$ for several different velocities $v$.  $\epsilon^{(2)} \propto 1/x^2$, and a different value of $x = \sqrt{x_1^2+x_\perp^2}$ was chosen for each velocity in order to make the qualitative effects easy to discern.}\label{ThetaDepend}
 \end{figure}

Combining \eqref{QSHighK} with the non-conserved contribution \eqref{TabDrag} we obtain for the energy density
 \eqn{CombinedT00}{
  \epsilon(t,\vec{x}) &\equiv 
   \langle T_{00}(t,\vec{x}) \rangle_{\rm tot} = 
   \int {d^3 k \over (2\pi)^3} e^{-ivk_1 t + i \vec{k} \cdot \vec{x}}
    \epsilon(\vec{k})  \cr
  \epsilon(\vec{k}) &= \epsilon^{(0)}(\vec{k}) + 
    \epsilon^{(2)}(\vec{k}) T^2 + {\cal O}(T^4)  \cr
  \epsilon^{(0)}(\vec{k}) &= 
   {v^2 \tilde{k}_1^2 - (2 + v^2) \tilde{k}^2 \over 
     24 \tilde{k}\sqrt{1-v^2}} \sqrt{g_{YM}^2 N}  \cr
  \epsilon^{(2)}(\vec{k}) &= i \pi v \tilde{k}_1 
    {2 v^2 \tilde{k}_1^2 + (5-11v^2) \tilde{k}^2 \over 
     18 \tilde{k}^4 (1-v^2)} \sqrt{g_{YM}^2 N} \,,
 }
where we have set $\tilde{k}_1^2 = (1-v^2) k_1^2$ and used $\tilde{k}^2 = \tilde{k}_1^2 + k_\perp^2$ as in \eno{kTildeDef}.  The integrals in \eqref{CombinedT00} can be easily computed by differentiating appropriately modified versions of the standard formulas
 \eqn{FourierTransf}{
  \int {d^3 k\over (2\pi)^3} {e^{i \vec{k} \cdot \vec{x}} \over \abs{\vec{k}}} = {1\over 2\pi^2 \abs{\vec{x}}^2} \qquad \int {d^3 k\over (2\pi)^3} {e^{i \vec{k} \cdot \vec{x}} \over \abs{\vec{k}}^4} = -{\abs{\vec{x}} \over 8\pi}\,.
 }
The energy density in position space can thus be written 
 \eqn{T00Position}{
  \epsilon(t,\vec{x}) &= 
    \epsilon^{(0)}(t,\vec{x}) + \epsilon^{(2)}(t,\vec{x}) T^2 +
      {\cal O}(T^4)  \cr
  \epsilon^{(0)}(t,\vec{x}) &= {(1-v^2)\tilde{x}_1^2 + (1+v^2) x_\perp^2 \over 12 \pi^2 (1-v^2) (\tilde{x}_1^2 + x_\perp^2)^3} \sqrt{g_{YM}^2 N}  \cr
  \epsilon^{(2)}(t,\vec{x}) &= -v \tilde{x}_1 {(5-11v^2) \tilde{x}_1^2 + (5-8 v^2) x_\perp^2 \over 72 (1-v^2)^{3/2} (\tilde{x}_1^2 + x_\perp^2)^{5/2}} \sqrt{g_{YM}^2 N} \,,
 }
where $\tilde{x}_1 = (x_1-v t)/(1-v^2)^{1/2}$ measures the distance along the $x^1$-axis in the co-moving frame, and $x_\perp = (x_2^2 + x_3^2)^{1/2}$.  At $t=0$ and for fixed $v$ and $\sqrt{g_{YM}^2 N}$, $\epsilon^{(0)}$ and $\epsilon^{(2)}$ are functions only of $\Delta\phi = \pi-\arctan x_\perp/x_1$, up to overall power-law dependences on $x = \sqrt{x_1^2+x_\perp^2}$.  In figure~\ref{ThetaDepend} we illustrate in two ways the effect of the ${\cal O}(T^2)$ term on the near-field energy density.

The main qualitative effects are Lorentz flattening of the region of large energy density, visible in figure~\ref{ThetaDepend}a, and a surprising pattern of energy build-up in front of the quark with a corresponding deficit of energy behind the quark, visible both in figure~\ref{ThetaDepend}a and in the $v=0.9$ curve in figure~\ref{ThetaDepend}b.  To understand this latter effect, recall that the subleading term $\epsilon^{(2)}$ encodes the leading behavior of dissipative effects at large $k$.  For $v \lsim 0.67$, this shift in the energy density is positive behind the quark (i.e.~for $\Delta\phi < \pi/2$).  This coincides with our intuition that the trailing string represents color flux spreading out behind the quark as it moves forward through the plasma.  But for $v \gsim 0.79$, the shift in the energy density reverses sign as a function of $\theta$.\footnote{In the intermediate range of velocities, $0.67 \lsim v \lsim 0.79$, the regions where $\epsilon^{(2)}$ is positive comprise a forward-pointing lobe and a backward-leaning, thickened cone.  This is exemplified in the $v=0.7$ curve in figure~\ref{ThetaDepend}b.}  A possible interpretation is that there is a ``bull-dozer effect'' where the quark's near field creates a pile-up of thermal gluons and other matter just ahead of it, like dirt piling up in front of the blade of a bull-dozer.  It would be interesting to study this effect further and see how far it survives beyond the leading correction term $\epsilon^{(2)}$.

\begin{figure}
  \includegraphics[width=6.5in]{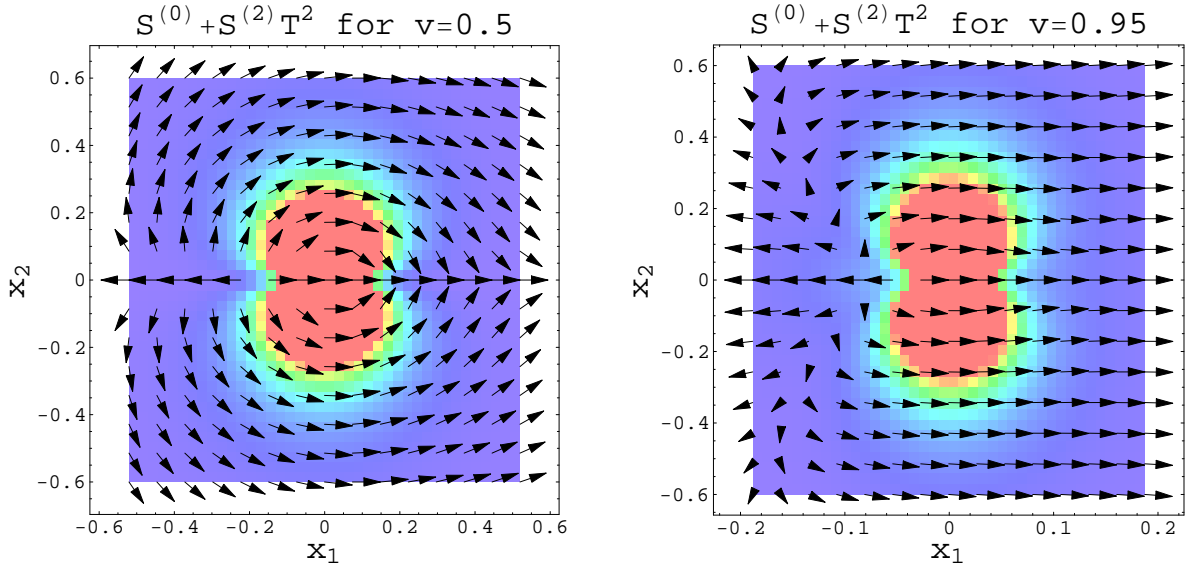}
  \caption{The near-field Poynting vector to the approximation shown in \eno{PoyntingPosition} in the plane $\pi T x_3 = 0.02$.  The arrows show the direction of the projection of $\vec{S}$ into the plane, and the color corresponds to the magnitude of of $\vec{S}$, where red means large and blue means small.}\label{Poynt}
 \end{figure}

Other components of the stress tensor may also be straightforwardly computed given $Q^{\rm tot}_S$, $Q^{\rm tot}_V$, $Q^{\rm tot}_T$, and the non-conserved terms \eno{TaiDrag}, \eno{TabDrag}, and~\eno{TijDrag}.  In particular, the Poynting vector is as follows:
 \eqn{PoyntingDef}{
  \vec{S} &= (S_1,S_2,S_3) \qquad
  S_i = S_i^{(0)} + S_i^{(2)} T^2 + {\cal O}(T^4)  \cr
  S_i(t,\vec{x}) &= -\langle T_{0i}(t,\vec{x}) \rangle_{\rm tot}
    = \int {d^3 k \over (2\pi)^3} 
      e^{-ivk_1 t + i \vec{k} \cdot \vec{x}} S_i(\vec{k})  \cr
  S_1(\vec{k}) &= -{v \over \sqrt{1-v^2}}
    {2\tilde{k}_1^2 + 3 k_\perp^2 \over 24 \tilde{k}} 
    \sqrt{g_{YM}^2 N} \left[ 1 + {v \over \sqrt{1-v^2}} 
     {8\pi i \tilde{k}_1 \over 3 \tilde{k}^3} T^2 + 
     {\cal O}(T^4) \right]  \cr
  S_2(\vec{k}) &= v {\tilde{k}_1 k_2 \over 24 \tilde{k}}
    \sqrt{g_{YM}^2 N} \left[ 1 +
      {v \over \sqrt{1-v^2}} {4\pi (7\tilde{k}^2 + 9 k_\perp^2)
        \over 3i\tilde{k}_1 \tilde{k}^3} T^2 + 
        {\cal O}(T^4) \right]  \cr
  S_3(\vec{k}) &= v {\tilde{k}_1 k_3 \over 24 \tilde{k}}
    \sqrt{g_{YM}^2 N} \left[ 1 +
      {v \over \sqrt{1-v^2}} {4\pi (7\tilde{k}^2 + 9 k_\perp^2)
        \over 3i\tilde{k}_1 \tilde{k}^3} T^2 + 
        {\cal O}(T^4) \right] \,.
 }
The Fourier integrals are again derivatives of the standard forms \eno{FourierTransf}, and the position-space forms are\footnote{We obtained the explicit position-space result \eno{PoyntingPosition} for the near-field Poynting vector after seeing a comparable result in a summary draft of \cite{YaromNew}.  We thank A.~Yarom for correspondence about this result.}
 \eqn{PoyntingPosition}{
  S_1(t,\vec{x}) &= 
   {v \over 1-v^2} {x_\perp^2 \over \tilde{x}^6} 
     {\sqrt{g_{YM}^2 N} \over 6\pi^2}
    \left[ 1 + {v \over \sqrt{1-v^2}} {\pi^2 \over 4}
      {\tilde{x}_1 \tilde{x} \over x_\perp^2}
      (2\tilde{x}_1^2+x_\perp^2) T^2 + 
      {\cal O}(T^4) \right]  \cr
  S_2(t,\vec{x}) &=
   -{v \over \sqrt{1-v^2}} {\tilde{x}_1 x_2 \over
     \tilde{x}^6} {\sqrt{g_{YM}^2 N} \over 6\pi^2}
    \left[ 1 - {v \over \sqrt{1-v^2}} {\pi^2 \over 12}
     {\tilde{x} \over \tilde{x}_1} (11\tilde{x}_1^2 + 8x_\perp^2)
      T^2 + {\cal O}(T^4) \right]  \cr
  S_3(t,\vec{x}) &=
   -{v \over \sqrt{1-v^2}} {\tilde{x}_1 x_3 \over
     \tilde{x}^6} {\sqrt{g_{YM}^2 N} \over 6\pi^2}
    \left[ 1 - {v \over \sqrt{1-v^2}} {\pi^2 \over 12}
     {\tilde{x} \over \tilde{x}_1} (11\tilde{x}_1^2 + 8x_\perp^2)
      T^2 + {\cal O}(T^4) \right] \,.
 }
In figure~\ref{Poynt} we show a cross-section of the approximation \eno{PoyntingPosition} Poynting vector field for $v=0.5$ and $v=0.95$.  For the purposes of plotting, a convenient way to regularize the divergence at $\vec{x}=0$ is to take a cross-section at fixed but small $x_3$.

\section{Conclusions}
\label{CONCLUSIONS}

We have derived fully decoupled forms of the Einstein equations linearized around a general five-dimensional warped background \eno{CurvedBackground} appropriate to describing holographic duals of finite temperature field theories on ${\bf R}^{3,1}$.  The final equations, \eno{MasterEQ}, \eno{MasterV}, and \eno{MasterS}, are expressed in terms of diffeomorphism-invariant master fields, from which one may extract all components of the metric perturbations.  Indeed, the entire derivation of the master equations has been cast in an explicitly diffeomorphism-invariant formalism.  Although our results for the scalar master equation are incomplete in that we did not give the full expression for the source term except in the $AdS_5$-Schwarzschild background, this general source term may be readily obtained starting from \eno{FullXYZ}.

We have applied the master field formalism to reproduce aspects of the results of \cite{Friess:2006fk}, which dealt with the stress tensor produced by an external quark moving through a thermal bath of ${\cal N}=4$ gauge theory.  We have also extended these results by calculating the leading correction at large wave-number to the Coulombic near-field of the moving quark.  This calculation was done using a method recently developed in \cite{Yarom:2007ap}.  The result for the energy density has an unexpected feature in position space: for $v \gsim 0.79$, the leading correction indicates a pile-up of energy just in front of the quark.

Following \cite{Friess:2006fk} we attempted to Fourier transform the numerical results to position space.  We found this difficult because even after the subtraction of the Coulombic near-field, the Fourier integrals have poor convergence properties: without the oscillating factor $e^{i \vec{k} \cdot \vec{x}}$, they are quadratically divergent in the ultraviolet.  With the leading correction subtracted off as well, this divergence is reduced to logarithmic, leading to a significantly more tractable numerical problem.  Ideally one should calculate to one more order of accuracy in the large $k$ asymptotic forms, so as to cure the divergences altogether and work with absolutely convergent Fourier integrals.  Obtaining the next correction is significantly harder than the calculations we did in section~\ref{TENSOR_LARGEK}--\ref{SCALAR_LARGEK}, but it may be tractable.

\section*{Acknowledgments}

This work was supported in part by the Department of Energy under Grant No.\ DE-FG02-91ER40671, and by the Sloan Foundation.  We gratefully acknowledge collaborative contributions by F.~Rocha in the early stages of this project, and we also thank J.~Friess and G.~Michalogiorgakis for useful discussions and A.~Yarom for correspondence.

\clearpage
\appendix

\section{Relation to the variables used in \texttt{hep-th/0607022}}
\label{JULYVARIABLES}

In \cite{Friess:2006fk}, the metric perturbations $h_{\mu\nu}$ were written as
 \eqn{JulyPerturbations}{
  h_{\mu\nu} = {L^6 \over r_H^3} \int {d^3 k \over (2\pi)^3} h_{\mu\nu}^{\vec{k}} e^{i \vec{k} \cdot \vec{x}}\,,
 }
where in axial gauge
 \eqn{hJulyDef}{
  h_{\mu\nu}^{\vec{k}} = \ell_v {r^2 \over L^3} \begin{pmatrix}
  H_{00} & 0 & H_{01} & H_{02} & H_{03}\\
  0 & 0 & 0 & 0 & 0\\
  H_{01} & 0 & H_{11} & H_{12} & H_{13}\\
  H_{02} & 0 & H_{12} & H_{22} & H_{23}\\
  H_{03} & 0 & H_{13} & H_{23} & H_{33}
  \end{pmatrix}\,.
 }
The equations of motion were decoupled into five sets of differential equations in quantities $A$, $B_1$, $B_2$, $C$, $D_1$, $D_2$, $E_1$, $E_2$, $E_3$, and $E_4$, which were defined as particular linear combinations of the $H_{mn}$'s (see \cite{Friess:2006fk}).  Setting $k_3 = 0$ and $k_2 = k_\perp$, and decomposing the above expression for $h_{\mu\nu}$ as in \eqref{hbcDecomp}--\eqref{hijDecomp}, we get
 \eqn{TVSToJuly}{\seqalign{\span\TL & \span\TR & \qquad\span\TL & \span\TR}{
  H_T^{T, {\rm even}} &= -\ell_v {L^5 v^2 \over 2 r_H^3 } {k_\perp^2 \over k^2} A &
  H_T^{T, {\rm odd}} &= -\ell_v {L^7 \over 2 r_H^4 } k C\cr
  H_T^{V, {\rm even}} &= -\ell_v {2 i L^5 v^2 \over r_H^3} {k_1 k_\perp \over k^2} D_2 &
  H_T^{V, {\rm odd}} &= -\ell_v {i L^9 v^2 \over r_H^5 } k_1 k B_2\cr
  f_t^{V, {\rm even}} &= \ell_v {2 L^4 r v \over r_H^3 } {k_\perp \over k} D_1 &
  f_t^{V, {\rm odd}} &= \ell_v {L^8 r v \over r_H^5 } k^2 B_1\cr
  H_L^S &= \ell_v {L^5 \over 3 r_H^3} E_3 &
  H_T^S &= \ell_v {L^5 \over 2 r_H^3 } E_4 \cr
  f_t^S &= \ell_v {2 i L^4 r v \over r_H^3} {k_1\over k} E_2 &
  f_{tt}^S &= \ell_v {2 L^3 r^2 h(r) \over 3 r_H^3} (E_3 - E_1)\,,
 }}
and of course $f_r^{V, {\rm even}} = f_r^{V, {\rm odd}} = f_{rr}^S = f_{tr}^S = 0$ in axial gauge.  We can use these formulas to relate $Q_T^{\rm tot}$, $Q_V^{\rm tot}$, and $Q_S^{\rm tot}$ to the quantities $Q^{\rm tot}_A$, $Q^{\rm tot}_D$, and $Q^{\rm tot}_E$, respectively, which were defined in \cite{Friess:2006fk}:
 \eqn{QRelations}{
  Q_T^{\rm tot} &= - {\ell_v\over \kappa_5^2} {r_H v^2 \kappa_\perp^2\over k^2} Q^{\rm tot}_A \qquad
  Q_V^{\rm tot} = {\ell_v\over \kappa_5^2} {4 k_\perp r_H v\over k} \left(Q^{\rm tot}_D + {i r_H\over 4 k_1 L^2 v} \right)\cr
  Q_S^{\rm tot} &= -{\ell_v\over \kappa_5^2} 2 r_H \left(Q^{\rm tot}_E + {i r_H v\over 2 k_1 L^2} \right)\,.
 }

\bibliographystyle{ssg}
\bibliography{source}

\end{document}